\documentclass[12pt,preprint]{aastex}

\makeatother

\newcommand{\js}{}
\newcommand{\etal}{{\js et~al.\/}}

\lefthead{Rigopoulou et al.~}

\slugcomment{to appear in ApJ}
\shorttitle{Spitzer Observation of Lyman Break Galaxies}
\shortauthors{Rigopoulou et al.}
\begin{document}
\title{SPITZER Observations of z$\sim$3 Lyman Break Galaxies: stellar masses
and mid-infrared properties}

\author{D. Rigopoulou,$\!$\altaffilmark{1}
J.-S.\ Huang,$\!$\altaffilmark{2}
C. Papovich,$\!$\altaffilmark{3 \footnote{Spitzer Fellow}}
M. L. N. Ashby,$\!$\altaffilmark{2}
P. Barmby,$\!$\altaffilmark{2}
C. Shu,$\!$\altaffilmark{4,5}
K. Bundy,$\!$\altaffilmark{6}
E. Egami,$\!$\altaffilmark{3}
G. Magdis,$\!$\altaffilmark{1}
H. Smith,$\!$\altaffilmark{2}
S. P. Willner,$\!$\altaffilmark{2} 
G. Wilson,$\!$\altaffilmark{7} and
G. G. Fazio$\!$\altaffilmark{1}
}
\altaffiltext{1}{Department of Astrophysics, Oxford University, Keble Rd, Oxford, OX1 3RH, U.K.}
\altaffiltext{2}{Harvard-Smithsonian Center for Astrophysics, 60 Garden Street,
Cambridge, MA 02138}
\altaffiltext{3}{Steward Observatory, University of Arizona, Tucson, AZ 85721}
\altaffiltext{4}{Joint Center for Astrophysics, Shanghai Normal University, Shanghai 200234, China}
\altaffiltext{5}{Shanghai Astronomical Observatory, Chinese Academy of Sciences, Shanghai 200030, China}
\altaffiltext{6}{California Institute of Technology, MS 105-24, 1201 E. 
California, Pasadena, CA 91125}
\altaffiltext{7}{Spitzer Science Center,
Caltech, 1200 E. California, Pasadena, CA 91125}

\begin{abstract}

We describe the spectral energy distributions (SEDs) of Lyman Break Galaxies 
(LBGs) at z$\sim$3 using deep mid-infrared and optical observations 
of the Extended Groth Strip, obtained with IRAC and MIPS on board Spitzer 
and from the ground, respectively.
We focus on LBGs
with detections at all four IRAC bands, in particular the 26 galaxies
with IRAC 8 $\mu$m
band (rest--frame K-band) detections.
We use stellar population synthesis models and probe the stellar
content of these galaxies. Based on best--fit continuous star-formation
models we derive estimates of the stellar
mass for these LBGs. As in previous studies, 
we find that a fraction of LBGs have very red
colors and large estimated stellar masses (M$_{*} > 5\times$ 10$^{10}$
M$_{\odot}$): the present Spitzer data allow us, for the first time, 
to study these massive LBGs in detail.
We discuss the link between these LBGs and submm-luminous galaxies.
We find that the number density of these massive LBGs
at high redshift is higher than predicted by current semi-analytic models 
of galaxy evolution.

\end{abstract}


\section{Introduction}

In recent years, great advances in our 
understanding of the nature and evolution of high redshift galaxies 
have been 
made, thanks to the availability of large samples. The techniques that have 
been developed rely either on colour selection criteria (e.g. Steidel et al.
2003, 2004, Franx et al. 2003, Daddi et al. 2004) or detection 
in the submillimeter through blank field surveys using the
Submillimeter Common User Bolometer Array (SCUBA) on the James Clerk
Maxwell Telescope (JCMT, e.g., Hughes et al., 1998) or the Max-Planck 
Millimeter Bolometer array (MAMBO, e.g. Bertoldi et al 2000, 
Eales et al. 2000). All these surveys have unveiled large numbers of $z>2$ 
galaxies but the overlap between the various populations is by no means
easy to investigate.

Among the various methods the Lyman break dropout technique 
(Steidel \& Hamilton 1993), sensitive to the presence of the 912$\AA$ 
break, is designed to select $z\sim 3$ galaxies.
The typical star formation rate (SFR) deduced from the UV
continuum emission of Lyman break Galaxies (LBGs) is estimated 
to be moderate, around 20--50 M$_{\odot}/$yr 
(assuming H$_{0}$=70 km s$^{-1}$Mpc$^{-1}$ and
q$_{0}$=0.5). The SFR value quoted is the ``mean'' SFR
which is derived from the entire LBG sample which ranges from 
10 to 1000 M$_{\odot}/$yr (for the most massive LBGs).
There is clear evidence, however, for the presence of
significant amounts of dust in LBGs (e.g. Sawicki \& Yee 1998,
Calzetti 2001, Takeuchi \& Ishii 2004 ). 
When corrected for the presence of dust then this ``mean'' SFR values 
increase to  
$\sim$ 100 M$_{\odot}/$yr.
The high SFRs and co-moving density of LBGs, together with the results of the 
clustering analyses (e.g. Adelberger et al. 2004) makes them plausible 
candidates for the long-sought progenitors
of present--day elliptical galaxies (e.g Pettini et al. 1998). 

The nature of the relationship between LBGs and submm-luminous
galaxies  (hereafter SMGs) has, however,  remained unclear. One
possible scenario is the one where LBGs and SMGs form a
continuum of objects with the submm galaxies representing  the
``reddest'' dustier LBGs.  Central to such a hypothesis is of course
the issue of dust in LBGs. A number of techniques have been used to
deduce the dust content of LBGs ranging from studies of the optical
line ratios (Pettini et al. 2001) to formal fits of the overall SED of
LBGs based on various star  forming scenaria (e.g. Shapley et
al. 2001, 2003).  Both methods agree that the most intrinsically
luminous LBGs, which have  higher SFRs, contain more dust (e.g. Adelberger 
\& Steidel 2000, Reddy et al. 2006). More
recently, X-ray stacking studies have shown that intrinsically more
luminous LBGs also contain more dust (e.g. Nandra 2002, Reddy \&
Steidel 2004). Another important issue in understanding LBG
evolution and their connection to SMGs is the study of their stellar
mass content.
The co-moving stellar mass density at any given redshift is the
integral of  past star-forming activity. Thus, stellar mass is a
robust tool to probe  galaxy evolution and is subject to fewer
uncertainties than the star formation rate (SFR).

Until recently, stellar mass estimates for z$\sim$3 LBGs have been based on
ground--based photometry which only samples out to rest--frame optical
band.  The observed luminosity at these wavelengths is dominated by
recent star-formation activity rather than the stellar population that
has accumulated over the galaxy's lifetime. With the advent of the
Spitzer Space Telescope (hereafter Spitzer,  Werner et al. 2004) we
now have access to longer wavelengths. In particular the availability
of the IRAC instrument (Fazio et al. 2004) with imaging  capabilities
out to 8 microns allow us to probe rest--frame K band (for  z$\sim$3
galaxies) where the light is sensitive to the bulk of the  stellar
content. Preliminary results of adding IRAC photometry to stellar mass
estimates have been presented in e.g. Barmby et al. (2004) for z$\sim$3 LBGs 
and Shapley et al. (2005) for BX$/$BM objects.

In this paper, we present sensitive mid--infrared photometry for a
sample of spectroscopically confirmed z$\sim$3 LBGs detected as part
of the IRAC Guaranteed Time Observations (GTO) program on the Extended 
Groth Strip (EGS).  We focus on
the broad band Spectral Energy Distributions (SEDs) of luminous LBGs
detected with  IRAC channel 4 (8 $\mu$m corresponding to rest--frame
K--band). The sample benefits from deep ground based optical
photometry and spectroscopy  (Steidel et al. 2003). Our aim is to
explore the range of stellar masses  and assess the benefits of adding
longer wavelength observations. We examine the correlation between
mass and absolute luminosity (magnitude) from optical to the K--band.
The paper is organised as follows: in Section 2 we present a brief
account of the observations and data reduction while in Section 3 we
review the sample properties followed by a discussion of the detailed
SEDs (Section 4).  We discuss the detailed models used to estimate
stellar masses in Section 5 while in Section 6 we present the actual
masses.  We focus on the properties of massive LBGs in Section 7 and
we discuss their possible evolution and connection to submm-luminous
galaxies. We estimate the number density of these massive LBGs in Section 8
and summarize our conclusions in Section 9.

\section{Spitzer Observations and Data Reduction}

The IRAC (Fazio et al. 2004) data presented here were acquired as part of the
IRAC GTO program.
The Extended Groth Strip was observed by IRAC 
at 3.6, 4.5, 5.8 and 8.0 $\mu$m during January and July
2004. IRAC observations of the EGS cover an area of 
2$^{\circ}$$\times$10$^{'}$. 
IRAC on board Spitzer  has the capability of observing
simultaneously in the 4 bands, when scanning a region, with an
effective F.O.V. of 5$^{'}\times 5^{'}$. The IRAC exposures consisted of 
52$\times$200 sec dithered exposures at each of the 3.8, 4.5, 5.8 $\mu$m 
wavelengths and 50$\times$208 sec exposures at 8$\mu$m, in the
2$^{\circ}$$\times$10$^{'}$ map. Because the field was observed twice and at
different position angles, removal of instrumental artifacts during
mosaicing was significantly facilitated.
The observations reach a (5$\sigma$) point-source sensitivity limit of 
24.0, 24, 21.9, and 22.0 mag (AB) at 3.6, 4.5, 5.8
and 8.0 $\mu$m, respectively.

Observations with MIPS (Rieke et al. 2004) were carried out in June 2004 in 
scanning mode. The MIPS 24$\mu$m channel ($\lambda=23.7 \mu$m;
$\Delta\lambda = 4.7 \mu$m) uses a 128$\times$128 BIB Si:As array with
a pixel scale of 2\farcs55 pixel$^{-1}$, providing a field of view of
5\farcm4$\times$5\farcm4. The scan map mode was used with the slow scan rate, 
which results in an integration time of 100 s pixel$^{-1}$ per scan pass 
(10 frames $\times$10 s) at 24 $\mu$m. Since the sky
position angle of the scan direction is determined by the spacecraft
roll angle, the observing dates were constrained such that the
resulting scan map would extend along the position angle of the Groth
Strip. The final scan maps cover a sky area of 2.4\degr$\times$10\arcmin\
with integration times of $>700$~s at 24 $\mu$m. 
The 24$\mu$m point source sensitivity (5$\sigma$) is 70 $\mu$Jy.

The IRAC and MIPS Basic Calibrated Data (BCD)
delivered by the Spitzer Science Center (SSC) include flat-field
corrections, dark subtraction, linearity correction and flux
calibration. The BCD data were further processed by each team's own
refinement routines. These additional reduction steps include
distortion corrections, pointing refinement, mosaicing and cosmic ray
removal by sigma-clipping. Source extraction was performed in the
same way as described in Ashby et al. (2006). In brief, we used
DAOPHOT to extract sources from both IRAC and MIPS images. With an FWHM of the
Point Spread Function (PSF) of 1$\farcs8$--2$\farcs0$ for IRAC and 6$\arcsec$ 
for MIPS 24 $\mu$m, virtually all objects at high redshifts are unresolved. 
We performed aperture photometry using a 3$\arcsec$ diameter apertures for
both IRAC and MIPS sources. The aperture fluxes in each band were 
subsequently corrected to total fluxes using known PSF growth curves 
from Fazio et al. 2004;
Huang et al. 2004. The magnitudes presented in this work are all 
in the AB magnitude system.

\section{The Spitzer LBG Sample}

The EGS LBG sample was constructed from the LBG catalogue of Steidel
et al. (2003). The catalogue was matched to the IRAC and MIPS source
lists instead of doing direct photometry on the images. We searched for
counterparts within a 2$^{''}$ diameter separation centered on the LBG
optical  position. The typical size for an LBGs is about 1'' and in most 
cases the LBG is clearly identified although
in Section 6.1 we discuss those cases where multiple counterparts were present.
Due to the resolution of 2'' we do not anticipate
finding many multiple component counterparts. 

Steidel et al. (2003) have reported on the detection of 334 LBGs in the EGS 
area. The Spitzer EGS survey covers 244 LBGs. About 200 LBGs are detected
with IRAC at 3.6$/$4.5 $\mu$m, 53 LBGs are detected with IRAC at 5.8 $\mu$m 
and 44 at 8.0 $\mu$m. Finally, 13 LBGs have counterparts in the
MIPS 24 $\mu$m survey of the field. 
Of the initial sample of 244 LBGs observed by Spitzer in the EGS area 175 
objects have confirmed
spectroscopic redshifts and are identified as galaxies at z$\sim$3 (we stress
that this number refers to objects classified as galaxies and that we exclude
from our analysis those classified as AGN, QSO or stars). 
Among
the 13 LBGs with MIPS detections, 6 objects are classified as galaxies at 
z$\sim$ 3 based on their optical spectra 
(of the remaining MIPS detections 3 are classified as QSO and 1 AGN). 

In this paper we focus on the properties of 44 LBGs that have
8~\micron\ detections (referred to hereafter as "the 8~\micron\ LBG
sample").  Of these, 26 have spectroscopic redshifts.  The
8~\micron\ LBG sample includes seven of the nine galaxies detected by
MIPS and in particular the Infrared Luminous LBGs (ILLBGs, Huang et
al. 2005).  The criterion for an LBG to be classified as ILLBG is
detection at 24~\micron, which for the sensitivity of the EGS survey
($\sim$70 $\mu$Jy (5$\sigma$) implies an infrared luminosity 
$L_{IR} > 10^{11}$ L$_{\odot}$.
The two LBGs with MIPS 24 $\mu$m detections but no 8 micron 
detection have not been added to the sample as their IRAC photometry is 
incomplete.

In Table 1 we list the ground based photometry for the 8 micron LBG sample. 
The UGRJK data and the spectroscopic redshifts
come from the large compilation of Steidel et al (2003). 
In Table 2 we list the Spitzer photometry.
The uncertainties in the IRAC $/$MIPS magnitudes were estimated from 
an analysis employing Monte-Carlo simulations. 
We added artificial sources to the images,
extracted them in the same manner as real sources, and computed the
dispersion between input and recovered magnitudes. The dispersion of
the recovered magnitudes about the mean recovered magnitude, 
computed for each bin in input magnitude, were
used as the uncertainty estimates.
In Figure 1 we plot the rest--frame UV$/$optical$/$near-infrared SEDs of 
all LBGs in the EGS area. The SEDs are normalised in R magnitude (R=24 mag).
The rest--frame UV (UGR) part of
the spectrum shows relatively little variation. Although K-band
magnitudes are available for $\sim$40\% of the EGS LBG sample we  
note a range of R--K colors with a mean value of $<$R--K$> \sim$ 1.5.

Clearly, the addition of the Spitzer IRAC$/$MIPS bands improves
dramatically our understanding of the nature of LBGs. The rest--frame
near-infrared colors of LBGs are spread over 4 magnitudes. 
LBGs display a variety of colors ranging from those that are red
with R-[3.6]$>$2 to blue with R-[3.6]$<$2. The SED of those LBGs with 
blue colors is rather
flat from the far--UV to the optical with marginal (in 1 band) IRAC detections.
On the other hand, the SEDs of the red LBGs are rising steeply towards longer
wavelengths. It is worth noting,
that a number of such red color LBGs display R-[3.6] values similar to the
extremely red objects discussed by Wilson et. al. (2004). 
Most of the ILLBGs display such extreme 
colors (see discussion in Section 7.2 and Figure 1 of Huang et al. 2005).  
The Spitzer observations presented here show that a fraction of 
optically-selected LBGs ($\sim$30\%) have red colors and have a
higher dust extinction and higher masses in contrast to the 
majority of  the LBG population which comprises mostly of objects with
blue colors modest extinction and masses.

\section{Population Synthesis Models}

\subsection{Model Parameters}

The SEDs of the Spitzer detected z$\sim$3 LBGs cover a wide range in
wavelength from $\sim$900$\AA$ to 2$\mu$m. In this Section we discuss the 
models used to fit
their SEDs in order to investigate and constrain the star formation
histories, extinction, and masses of the LBGs.  
We use the Bruzual \& Charlot code (2003, hereafter BC03) to 
generate models in order to fit the LBG SEDs.
The new BC03
models are based  on a new library of stellar spectra with an updated
prescription of AGB stars. As suggested by the authors we adopted the
Padova 1994  stellar evolution tracks and constructed models with
solar  metallicity (see discussion in Shapley et al. 2004) and  a 
Salpeter Intitial Mass  Function (IMF) extending 
from 0.1 to 100 M$_{\odot}$. We use the
Calzetti et al. (2000) starburst attenuation law to simulate
the extinction. 

A major uncertainty in this type of analysis is the parameterization
of  the star formation history (e.g. Papovich et al. 2001, Shapley et
al. 2001, Bundy et al, 2005). For the present analysis we have 
considered mostly two
simple single-component models: exponentially declining models of the
form  SFR(t) $\propto$ exp(-t$/ \tau$) with e-folding times of $\tau$
= 0.05, 0.1, 0.5, 2.0, and 5.0 Gyr and, continuous star formation
(CSF) models. 
Although complex models, such as combinations of various
star forming histories or single bursts on top of a maximally old 
underlying bursts (as suggested by Papovich et al. 2001) are probably more
realistic, we do not consider them in this work as we cannot constrain the 
model parameters easily. 

Our main aim for each galaxy is to constrain the stellar population
parameters. The fitted parameters are the following: dust extinction
(parameterised by E(B--V)), age (t$_{sf}$ defining the onset of star formation), stellar Mass (M$_{*}$) and
star formation history ($\tau$). Using BC03 we generate a grid of models
with ages ranging between 1 Myr and the age of the Universe at the
redshift of the galaxy in question and, varying extinction. For each
set of E(B--V), star formation history, and age we derive a model with
the full set of colors, UGRJK plus the IRAC 3.6, 4.5, 5.8 and 8 $\mu$m, placed
at the redshift of the galaxy in question. The intrinsic model spectrum 
is then reddened by dust. The model SED is finally corrected for the 
intergalactic medium (IGM) opacity.
The predicted colors are then compared with
the observed ones using $\chi^{2}$ minimization technique. 
The best-fit E(B--V) and age combination was chosen to minimize $\chi^{2}$
and the intrinsic SFR and stellar mass were determined from the normalization 
of the best-fit model to the measured SED. 

We estimated the confidence intervals for individual objects using
Monte  Carlo simulations. For each source we generated about 200
synthetic model SEDs of the data by varying the fluxes randomly
(random values where chosen according the the Gaussian distribution of
the measured uncertainties). We then repeated the fitting procedure
for each new set and determined the 68\% confidence intervals from the
distributions of the best-fit values obtained with each set. The
procedure used here is similar to the one used recently 
by Papovich et al. (2006) for DRGs, Shapley et al. (2001) and
Papovich et al. (2001) for LBGs. We note a strong assymetry around
best-fit values which we interpret as the presence of strong
degeneracies especially between E(B--V) and t$_{sf}$ (as we discuss in
Section 5.2). We note however that, the inferred stellar mass is one of the 
best constrained quantities suffering much less from uncertainties involved
with the specific star formation history used to describe a specific LBG.
In what follows we discuss how varying properties of the model fits affect 
the best-fit SED parameters.

\subsection{Extinction}

The impact of different extinction laws has already been investigated
by e.g. Papovich et al (2001), Dickinson et al. (2003) who found the
effect to be overall small.  For the present work we have adopted the
Calzetti law since such a law reproduces the total SFR from the
observed UV for the vast majority of LBGs (e.g. Reddy \& Steidel 2004,
Reddy et al., 2005, Nandra et al. 2003). The choice of the Calzetti
law was also dictated by the desire to facilitate comparison with
previous work in the field.

\subsection{Initial Mass Function}

As discussed in 4.1 we have chosen a Salpeter IMF for our stellar
population models. In this Section we investigate the effect on the
models when using different IMFs. A Scalo (Scalo 1986) or a
Miller-Scalo (Miller \& Scalo 1979) IMF are also consistent with the data, 
as long as we keep
the model lower and upper mass cutoffs  fixed at 0.1 and 100
M$_{\odot}$, respectively.  Both the Scalo and the Miller-Scalo IMFs
however, result in slightly younger  ages than the ones obtained here
using Salpeter IMF. A Chabrier IMF (Chabrier 2003), also
behaves very
similarly to the Salpeter IMF. Both of them have a very similar
upper mass dependence. However, in the Chabrier IMF the low-mass
end assumes a flatter behaviour, following a
log-normal distribution. This IMF results in M$/$L ratios that are a 
factor of $\sim$1.5 smaller than those for a Salpeter IMF.

Overall the exact values of the upper$/$lower
mass cutoffs have a more noticeable effect on the derived stellar
masses. The major difference lies in the relative contribution of stars 
with M$<$M$_{\odot}$. A lower mass cutoff of 1M${_\odot}$ 
would, for instance, result in stellar mass values  that are lower by 
about 30--40\% than those derived when the lower mass cutoff is set 
to 0.1M$_{\odot}$ (assuming a Salpeter IMF). The result would be more 
pronounced (i.e. the estimated stellar mass
will be lower) if we use a Chabrier (2003) IMF instead. 
Bell et al. (2003) seem to favour a ``diet'' Salpeter IMF (a Salpeter 
IMF which is deficient in low mass stars) which in the local Universe 
provides a satisfactory explanation for the M$/$L ratios derived from colors. 
Since the stellar content of $z\sim3$ galaxies and in particular the fraction
of low-mass stars is presently unknown we do not investigate variations in 
the lower mass cutoff any further.

\subsection{Metallicity}

So far, information on element abundances in LBGs is rather limited. 
Pettini et al. (2002) determined element abundances in cB58, 
a typical L$_{*}$ galaxy which benefits from a factor of 30 magnification,
and found it to be $\sim$ 0.25 Z$_{\odot}$. 
Nagamine et al. (2001) suggested that near--solar metallicities are in fact
common in z$\sim$3 galaxies with masses greater than 10$^{10}$ M$_{\odot}$ 
which is broadly consistent with the results for cB58.
Shapley et al. (2004) also argued for solar metallicities for z$\sim$3 LBGs.
Because 8~\micron-detected
LBGs are likely to be more massive than the typical LBG at $z=3$,
we used solar metallicity in the models.  Reducing metallicity to
half solar would decrease the derived masses by 10--20\%.


\subsection{Model Uncertainties}

Besides photometric uncertainties the parameters derived from SED fitting 
suffer from systematics and are subject to degeneracies simply because
the models cannot fully constrain the star formation history of a high 
redshift galaxy. The uncertainties are model-independent and plague even the
simplest single-component models that we use in the present work. This
is due to the fact that model parameters such as extinction and star formation 
are strongly dependent on the value of $\tau$ used to parameterise the star
formation history. 

The extinction E(B-V) is strongly dependent on $\tau$ and the exact
prescription used to describe the star formation history. For example
a decaying star formation model with a lower E(B-V) will produce the
same G-R and R-K colors as a model with a larger $\tau$, smaller
$\tau /$t$_{sf}$ but a  higher E(B-V). For constant star formation
histories extinction is responsible for reddening the UV part of the
spectrum which is made up of a mixture of stars of earlier types.
Likewise, the inferred age, t$_{sf}$, for a given value of $\tau$
depends on the strength of the Balmer break. The latter is sensitive
to the relative number of B, A and F stars with 
respect to late type stars and the CaII H and K absorption 
at ~4000 $\AA$ (which is determined by the relative number of late 
type stars). 
A larger value of $\tau$
corresponds to older ages for the O and A stellar populations. Finally,
the SFR also suffers from uncertainties since it is primarily derived
from the UV slope which will correspond to
different E(B-V) and dust attenuation factor depending on the assumed
stellar population and extinction law.

The uncertainties we just discussed are of course likely to affect the
inferred stellar masses but in a less dramatic way. Even in the case
of single star formation histories (CSF or EXP) the
derived stellar masses vary by  a factor of $\sim$10\% for different
values of $\tau$. Uncertainties in extinction and age also
influence the derived masses but, it is quite difficult to
disentangle the effect each of these have on the masses.  A highly
extincted underlying stellar population will have a similar effect on
the stellar mass as an older population. As we show in Section 5,  
the
addition of rest-frame near-infrared photometry from Spitzer has
helped to constrain the dust properties of 8 micron selected LBGs.

Uncertainties in the mass estimates become more serious
when one assumes more complex star formation histories. Papovich et
al. (2001), for instance, introduced underlying maximally old bursts
with t$_{sf} >>$$\tau$ and found an increase in the M$/$L by a factor
of several without a noticeable effect on the UV colors. Glazebrook
et al. (2004) introduced random bursts in the range of star formation
histories used to fit their multiband photometry and found that the
mass increased by about a factor of 2 although the amount by which the
stellar mass is underestimated depends on the specific galaxy SED.  We
conclude that use of the simple star formation histories (CSF or EXP)
is likely to provide a lower limit on the stellar mass estimate.

Finally, it is worth noting that stellar population models including the 
Thermally-Pulsating Asymptotic Giant Branch (TP-AGB) phase (Maraston 2005)
can also be used to fit the SEDs of z$\sim$3 galaxies. The signature of the
TP-AGB phase is quite prominent for ages 0.2--2 Gyrs and as van der Wel et al.
(2005) points out they provide a better fit for the SEDs of 
their z$\sim$1 galaxies. We defer modelling of our z$\sim$3 LBGs with 
models including the TP-AGB phase for future work.


\section{Stellar Masses}

Using the stellar population synthesis models described in Section 5.1 we have
estimated the stellar masses for the entire sample of 181 LBGs with confirmed
spectroscopic redshifts. Figure 2 shows a histogram of the inferred stellar 
masses for the whole Spitzer LBG sample.
From this sample we selected LBGs 
with 8 micron 
detections and carried out a detailed study of their properties. 
In Figure 3 we show the rest--frame UV--optical--near-IR SEDs
together with best-fit models for the 8 micron LBG sample (26 objects
with confirmed spectroscopic redshifts). In Table 3 we report the
results from the 
best-fit stellar population models for each galaxy for the entire
optical--IRAC SED. Parameters are listed for the best-fit model
either CSF or exponentially decaying models, whichever gives the lowest
value of $\chi^{2}$ with $\tau$ as a varying parameter.
The stellar mass inferred from
the best fit model along with the fit parameters for each object are reported
in Table 3.
The median age for our sample is t$_{sf}$ = 700 Myr. More than 50\% of our
galaxies have t$_{sf} >$ 500 Myr while only 15\% have t$_{sf} <$ 100 Myr. 

As we discussed already in Section 3, the 8 micron LBG sample contains most
(all but two) of the ILLBGs (Huang et al. 2005). ILLBGs appear to have
red colors (R-[3.6]$>$3) and best-fit ages t$_{sf} > $1000Myr. 
Using best-fit CSF models we compute the stellar masses of the ILLBGs 
and find them to be of the order of (M$>$10$^{11}$M$_{\odot}$). Extinction for
these massive LBGs (parameterised by E(B--V)) is around 0.3.
The fits of the ILLBGs are discussed in detail in Section 7.1.
The inferred stellar masses quoted in Table 2 do of course suffer from the 
uncertainties discussed in Section 4.6 and are quantified with the error bars
reported in the same Table. The median stellar mass for the 
8 micron LBG sample is (8.16$\pm$1.04)$\times$10$^{10}$ M$_{\odot}$ 
compared 
to (2.95$\pm$1.51)$\times$10$^{10}$ M$_{\odot}$ estimated for the entire 
EGS-LBG sample (and excluding the 8 micron selected LBGs). 
The fraction of LBGs with
M$>$5$\times$10$^{10}$M$_{\odot}$ among the 8 micron LBGs is 40\% compared to 
25\% for the entire EGS-LBG sample.

Finally, it is also worth investigating the morphology of the 8 micron LBG
sample. This investigation was carried out to assess at what level
the IRAC fluxes might suffer from contamination from neighbouring sources.
This is important as the resulting fluxes affect the modelling and the 
computation of stellar masses.
For this reason we have compared the 8 micron IRAC images with deep
R--band SUBARU images (Miyazaki,  priv. comm). As we discussed in Section 2 the
IRAC PSF is $\sim$2$^{''}$ so we are looking for multiple sources within
a 2$^{''}$ area. Out of the 26 sources in the 8 micron LBG sample we have 
identified
two cases where a possible contamination of the IRAC 8 micron flux
by a neighbouring source is possible. In Figure 4 we show cutouts of
the optical R-band images for Westphal-M38 and Westphal-MD99. 
While we include these sources in the discussion we
caution that the reported fluxes might be contaminated.

\subsection{Dust extinction: are 8 micron LBGs dustier?}

In this Section we try to investigate the dust content of the 8 micron LBG 
sample based on the E(B--V) values derived from the best-fit models.
In Figure 5 we plot the estimated stellar mass (based on best-fit CSF models) 
as a function of best-fit E(B--V) for the 8 micron LBG sample. 
We chose to plot E(B--V) against the stellar mass which, as we discussed
extensively in Section 4, is relatively insensitive to the assumed star 
formation history.
For comparison we also plot the same parameters for the LBG-EGS sample 
(i.e. all LBGs with detections in at least one IRAC band). 
The mean E(B--V) for the LBG-EGS, the 8 micron LBG and the ILLBG samples
IS 0.156, 0.232 and 0.354, respectively. 
The more massive LBGs have, on average, higher E(B-V) though
with a considerable scatter. 
We confirm that 8 micron LBGs suffer higher extinction 
(parameterised by the model value of E(B--V)) when compared to the
Spitzer LBG sample (ie those without 8 micron detections). 
Moreover, extinction is higher for the ILLBGs
although our sample is at present still small (6 objects) to draw
statistically significant conclusions. 

Another way to investigate the amount of obscuration in LBGs is 
to look at the 24 micron detections (ie the ILLBGs). Out of the
188 LBGs in the LBG-EGS sample only six of them are detected 
at 24 microns that is a fraction of $\sim$3\%.
However, the percentage of 24 micron detections among the 8 micron LBG sample
is higher -- 6 out of the 26 LBGs-- which makes up for 23\%.  
We thus conclude that Spitzer observations provide reliable means to probe the
dust content of LBGs. In particular objects with Spitzer 
24 micron detections
must be in the high extinction end of the dust distribution.
 

\section{Rest--frame NIR properties of LBGs}

The addition of longer wavelength photometric points (ie the IRAC
channels) has an effect on the accuracy of stellar mass
estimates (e.g. Labbe et al. 2005, see also discussion in Section 7)
but also on our understanding of the properties of the LBG population as a 
whole.  
Here, we investigate the distribution of stellar mass with
rest--frame wavelength as we move from optical to the near-infrared bands.
In Figure 6a we show the distribution of
stellar masses as a function of absolute [3.6] $\mu$m magnitude which
at the median redshift of 3 corresponds to rest--frame 0.9 $\mu$m (ie
I--band). Although there is clearly a correlation between absolute
I-band magnitude and stellar masses (correlation coefficient r=0.53) 
there is considerable scatter in
the values especially at the fainter end. At the brighter end (which is where
most of the massive galaxies are found) 
the correlation becomes tighter but this might also be due to the smaller
number of sources at these magnitudes. We suggest that the spread in the 
correlation is due to the wide range of star--formation histories among LBGs.
Moreover, it is apparent that it is quite difficult to project a single 
I-band rest--frame luminosity to a stellar mass.

In Figure 6b we show the distribution of stellar mass as
a function of absolute [5.8] $\mu$m magnitude which at z=3 corresponds to
rest--frame  1.4 $\mu$m (ie H--band). The correlation between stellar mass
and absolute rest--frame H-magnitude improves significantly (r=0.61).
The scatter in stellar mass decreases
as one moves to longer
wavelengths and probes the stellar luminosity due to recent$/$slightly older
 star formation activity. Likewise, in Figure 6c we plot the distribution of
stellar masses as a function of absolute [8.0] $\mu$m magnitude which would
correspond to rest--frame $\sim$2.0 $\mu$m (i.e. K--band). The correlation
between stellar mass and magnitude becomes even tighter with r=0.77. The 
scatter in M$/$L values decreases and is now a factor of $\sim$12. The values
of M$/$L for the most massive galaxies show a spread of about 10. 
From these simple comparisons, we conclude that the 
mid--infrared bands and especially the IRAC 8 $\mu$m channel 
(which samples the rest frame near--infrared wavelengths
for z$\sim$3 LBGs) provides a more accurate estimate of the M$/$L ratio 
compared to that obtained when using optical bands. As we discussed already, 
 IRAC 8 $\micron$ channel is sensitive to the light from the bulk of 
the stellar activity accumulated over the galaxy's lifetime (see also Bell
\& de Jong 2001). 

We finally examine the dependence of stellar mass on the R--[3.6] color index. 
As we discussed already in Section 5,
the massive LBGs tend to have  red colors with R--[3.6]$>$2. In Figure
6d we plot stellar  masses as a function of the R--[3.6] color.  The
entire EGS--LBG sample (with confirmed spectroscopic redshifts) shows a
wide range of R--[3.6] color with the 8 micron LBG members showing
colors in the range 1 to 4. The most massive 8 micron LBGs (which are
also ILLBGs) show the most extreme R--[3.6] colors $>$2.5. These
values are close to  the values of extremely red objects presented by
Wilson et al. (2004). 
The correlation is tighter for the more
massive$/$redder LBGs which have M$/$L close to that of present day
galaxies. 

All these results stress the importance of obtaining longer wavelength 
(rest--frame near-infrared) observations for the z=3 LBGs,
as these allow us to probe the 
effect of recent star formation on the galaxy luminosities and masses.
Although proper knowledge of the current stellar content is crucial, 
as we have discussed, it is the luminosity of older stars, 
that dominates the K--band light, that is found to correlate best with the
derived stellar mass. In particular, this correlation becomes significant 
for the
newly discovered ILLBGs and the massive 8 micron sample LBGs which have a 
significant dominant older stellar population. As we move to the other 
extreme of
young and therefore less massive galaxies the current stellar content becomes
perhaps more important as it dominates the luminosity and thus the stellar
mass.

All the above correlations have been based on the stellar
masses inferred from the models described in Section 3. To some extent
the results are dependent on the particular details of the models
assumed, in our case a single exponentially decaying or continuous
episodes of star formation. Our models are, however, rather
``conservative''  in terms of the predicted stellar masses. 
If we use a two component model  (an underlying old stellar
episode and a very young continuous episode) this  will result in a
relative increase in the derived stellar masses which will be  higher
for the lower mass galaxies in comparison to the higher mass galaxies
(Papovich et al. 2001, Shapley et al. 2005).
This will likely make the trends seen in Figure 6 shallower but will
not significantly affect our conclusions.

\section{Comparison with K-selected LBGs}

The original selection of LBGs was based purely on optical colors 
(Steidel \& Hamilton 1992) thus only probing the rest--frame UV part of their 
spectrum. Shapley et al. (2001) and Papovich et al. (2001) obtained 
near-infrared photometry thus extending the available SEDs into the rest--frame
optical. With the present data we have extended the SEDs of LBGs into the 
rest--frame near-infrared and have the opportunity to
probe the bulk of the stellar emission. As we discussed already in Section 3 
about 1$/$4 of the EGS-LBG sample has been detected with IRAC [8.0] 
(that probes the rest--frame K--band). In this Section, we  
want to compare the properties
of the 8 micron LBG sample to those of optical$/$near infrared selected 
samples such as the NIRC sample (Shapley et al. 2001).

There are 16 LBGs in common between the NIRC (Shapley et al. 2001) and the
entire LBG-EGS  sample while five of these are part of the 8 micron 
LBG sample. Of these five targets we chose Westphal-D49 to carry out 
a detailed comparison. 
The best-fit CSF model for D49 (Table 3) has an age of t$_{sf}$=1350
Myr and an E(B--V) = 0.34. In the absence of the Spitzer photometry
Shapley et al. fit D49 with a CSF model with 1139 Myrs and E(B--V)=0.17
In Figure 7 we want to compare the two models. 
It is obvious that in the absence of the Spitzer
data both models would provide a good fit to the rest--frame
UV--optical photometry. It is only with the addition of the
rest--frame near infrared (Spitzer data) that a model with higher
extinction becomes necessary. One could of course argue for a
more ``aged'' and less extincted stellar model. Likewise, we have
generated models where we varied the age of the stellar population
(Figure 8). It is obvious that stellar ages larger than 1.3 Gyrs do not
provide a satisfactory fit to the SED.

A similar exercise was performed for the remaining objects that are in common
between the NIRC and our 8 micron LBG sample. For objects without
8 micron detections (11 LBGs) we find a very good
agreement in the models (in particular age and extinction). The mean age
of the NIRC sample is 355 Myrs compared to a mean age of 320 Myrs derived 
including the IRAC detections. Likewise the mean extinction derived for
the NIRC sample is 0.218 whereas our mean extinction is 0.208. The differences
are more pronounced however, for LBGs with 8 micron detections. Out of the 
small sample of 5 LBGs we derive mean age and extinction of 400 Myrs and
E(B-V)=0.378, while the corresponding values for the NIRC fits are 293 
Myrs and 0.238.
For two of the five targets in common between NIRC and 8 micron
samples (MD109 and MD115) Shapley et al. could not constrain the best-fit ages
and thus fixed the age at 10 Myrs. Figure 9 summarizes our results.

The overlap between the NIRC sample and the 8 micron LBG sample is small
so the differences might not be very significant. 
We suggest, however, that it is likely that LBGs with
8 micron detections have higher extinctions (see also discussion in 5.1) and
slightly more aged population than LBGs undetected at 8 micron. 
The possibility for 
8 micron contributions from a putative AGN cannot be constrained with 
the current data although optical spectroscopy (Steidel et al. 2003) or
X-ray studies (e.g. Nandra et al. 2002) do not reveal any signs of such 
activity.

\subsection{Massive LBGs and ILLBGs}

A subset of the 8 micron LBG sample (12 objects out of 44) have masses
M$_{*} \geq$10 $^{11}$ M$_{\odot}$. From this subset 1 object is a
confirmed  QSO and six objects do not have spectroscopic
confirmation. In Section 5 we discussed in details the various
models used and the accuracy in the derived stellar masses. In this
Section we focus the discussion on these ``massive'' 8 micron LBG
sample  galaxies with confirmed spectroscopic redshifts. These massive
8 micron LBGs are also detected in the MIPS24 $\mu$m band. There are
two additional galaxies that do not have 8 micron counterparts but are
detected in the 24 MIPS band, we will consider those as part of the
massive luminous LBG sample.

Because of their luminosity and rather unusual properties (compared to
the rest of the LBGs) Huang et al. (2005) termed this subclass of
objects as Infrared Luminous LBGs (ILLBGs).  The masses of the
ILLBGs have been estimated using BC03 models and are reported in
Table 2. All ILLBGs are
best fit by a CSF model with ages $>$ 1000 Myr with extinction that varies
around  A$_{V} \sim $0.2. 
Although, as we discussed, the
addition of the IRAC bands has significantly improved the accuracy of
the derived stellar masses, the models do not constrain
very well the star formation history of high-z and LBG objects (as has been 
discussed already by Papovich et al. 2001, Shapley et al. 2004, 2005). 
Despite the uncertainties in the detailed properties of the star
formation history the derived stellar mass is a robust estimate
depending mostly on the SED shape rather than the details of the
particular model.  However, we will show that in the case of these
``massive'' luminous LBGs we  can, for the first time, discriminate
between the two ``simple'' models (CSF vs EXP) used to fit the SEDs of
the LBGs. We caution that although the star formation history of
galaxies is a more complicated process than a constant star formation
activity or a a smooth exponential function the details of more
complicated models (short bursts of different strength  superimposed on a
constant star forming model) can not be well constrained so we do not
consider them here.

As we have shown in Section 6 the correlation between estimated stellar
mass and absolute magnitudes becomes tighter as one moves to longer
wavelengths especially for the brighter more massive objects. An
accurate estimate of the total stellar mass can be obtained when the
star formation of the past several hundred Myrs is less than the
integrated star formation over the lifetime of a galaxy. In a more
realistic case, a good estimate of the stellar mass can be obtained
when the light from the optical and more importantly near-infrared
wavelengths far exceeds the UV light. The massive ILLBGs considered here
represent exactly this phase in galaxy evolution: their light is dominated
by the light of the older stars which output mainly in the near-infrared bands
(observed frame around 8 $\mu$m covered by IRAC [8.0] channel) and exceed 
the light in the UV (due to the more
recently formed stars). This is in line with the large M$/$L$_{2.2}$ found 
for the massive LBGs.

The last point to be addressed is the issue of the exact prescription
of star formation history used to estimate masses for the ILLBGs.
The discussion in Section 6 is of particular relevance. There, we discussed 
how the
luminous (and massive) LBGs show the most extreme (very red) R-[3.6] 
and g-R colors. Through our detailed modelling we 
find that CSF models with an  
extinction E(B--V)$\sim$0.3, reproduce quite well the extreme colors of 
ILLBGs (see also Figure 2 in Huang et al. 2005). In models with 
declining star formation histories red G-R and R--[3.6] colous
can be produced either with t$_{sf} / \tau > $1 and little or no dust 
extinction or with smaller t$_{sf} / \tau$ and more dust 
extinction. In such models the reddening of the UV slope is caused by 
a mixture of stars of later spectral types.
In continuous star formation models reddening of the rest-frame
UV is due solely to extinction. The fact that ILLBGs are detected at 
24$\mu$m implies that they are quite dusty so that the constant star 
formation model provides a natural match to the extremely red colors observed.
Since (observationally) we cannot constrain the spectral types of late type
stars in LBGs we suggest that the CSF model is a better (simpler) explanation
for the extremely red colors displayed by ILLBGs.

\subsection {\bf {Old and Massive LBGs: a new class or the missing link?}}

We now want to investigate the connection between the massive luminous
z$\sim$3 LBGs and the SMGs. Both LBGs and SMGs have sufficient SFRs to
form present-day elliptical galaxies. Until recently, attempts to find a
connection between the two populations had not been very successful
(Webb et al. 2003,  Chapman et al. 2000) although more recent studies (e.g.
Chapman et al. 2005, Reddy et al 2005) have been exploring a possible link.
Naturally, most of the submm 
searches targeted LBGs
with high predicted SFRs. These SFRs, however, have been estimated
based on the relationship between far-infrared$/$far-UV flux and UV
continuum for local starbursts (Adelberger \& Steidel 2000); this
relationship however, does not apply to ultraluminous galaxies
where a greater fraction of the total star formation is in
optically-thick regions.
Another issue is of course the detailed 
properties of the dust distribution in LBGs which may differ from that of 
present-day evolved galaxies (Takeuchi 2003). 
An obvious evolutionary scenario is one in which 
massive LBGs and SMGs form a continuum of objects with
the bright submillimeter-selected sources representing the highest
star-forming LBGs. 

The current sample of massive LBGs has many properties in
common with SMGs. For one, as we discussed in Section 6
their R--K colors are extreme ($>$3) similar to those of SMGs. 
Indeed the only  LBG detected in the submm so far , Westphal
-MMD11, by Chapman et al. (2000) shows such extremely red colors. The
Spitzer colors of massive LBGs are also very similar to those of
SMGs. Egami et al. (2004) reported on Spitzer detections of SMGs
in the Lockman hole. Based on the observed SEDs of the SMGs
and from comparisons with local templates they classified SMGs
as warm (Mrk231-like SED) and cold (Arp220-like SED). They
found that  the majority of SMGs are cold. More recently, Ashby et al. (2006) 
presented a detailed study of the mid-colors of SMGs detected
in the EGS area. Using these
measurements as a reference point, we plot in Figure 10 (a and b) the R-[3.6] 
and [8.0]-[24.0] colors for SMGs and ILLBGs and find that they are indeed 
quite similar. For the SMGs we used the 17 candidates with
secure 8 $\mu$ counterparts (see note in Table 1 of Ashby et al. 2006). 
Finally, it is worth looking at the inferred stellar masses of SMGs and LBGs:
both populations appear to have large estimated stellar masses ranging
from several $\times$10$^{10}$ to 10$^{11}$ M$_{\odot}$ (e.g. Borys et al.
2005, Tecza et al. 2004). We caution that 
stellar mass estimates for SMGs are relatively uncertain due to the likely 
presence of an AGN component in some members. The values we quote are based
on the estimates by Borys et al. (2005) for relatively young and gas-rich SMGs
as well as the dynamical masses measured by Tecza et al.
Based on the similarities of the colour indices, mid-infrared colors 
and inferred stellar masses we propose that a link must exist between
ILLBGs and SMGs.

To further investigate the possible connection between the two populations let
us also take a look at some sample statistics. Huang et al. (2005)
found that ILLBGs represent about 15\% of the total LBG
population, excluding objects with possible contamination from AGN. This number
most likely represents a lower limit to the true fraction of ILLBGs
among the general LBG population for two reasons:(a) the MIPS depth is
not uniform over all of the Steidel et al. (2003) fields  and (b) we
do not have spectroscopic confirmations for the whole ILLBG sample.

Recently, Chapman et al. (2005) presented spectroscopy of 73 SMGs with faint 
radio emission. The median redshift of the Chapman et al. sample is 2.2 with 
19 objects with redshifts z$_{spec} >$2.5. Although the redshift 
distribution of the SMGs does not match that of the z$\sim$3 LBGs we will
use the Chapman sample as is the largest compilation of measured spectroscopic
redshifts for SMGs (though we have to make some adjustments to allow
for the slight difference in the mean redshifts) for our comparisons. 
Chapman et al. find that 30 of the 73 SMGs with confirmed spectroscopic
redshifts have rest-frame UV-characteristics similar to those of star 
forming galaxies. Among the 19 z$>$2.5 SMGs, 6 of them have starburst-like UV
colors (similar to those of LBGs). 
Thus, we estimate that $\sim$30\% of SMGs are star-forming
galaxies (with UV colors matching those of LBGs) with this number likely 
to be a lower limit  (due to
unidentified objects which are unlikely to show AGN signatures). 
Among the EGS-LBG sample, on the other hand, Huang et al. found that 15\% 
of LBGs (the so-called ILLBGs) show 
evidence for increased dust extinction (via their 24 $\mu$m detections).
Very recently we confirmed detection of ILLBGs at 
mm wavelengths with IRAM$/$MAMBO (Rigopoulou et al. 2006, in prep) 
at a flux level similar to that of faint SMGs.

Bringing these statistics together 30\% of SMGs display LBG-like
UV-colors while 15\% of LBGs show evidence for higher extinction and submm 
emission. Thus, it is likely that ILLBGs (with red colors,
higher A$_{V}$, submm emission) and a sub-fraction of SMGs 
(those with UV colors
similar to starburst galaxies) appear to be identical in their 
UV and$/$or submm properties, this is likely to happen at the 2--3 mJy level.
It is possible that SMGs fainter than the current survey limits (2-3 mJy) 
will have lower star formation rates and lower levels of dust obscuration, 
such that the overlap fraction between SMGs and LBGs would become increasingly
larger as one goes to fainter flux levels.
We suggest that an evolutionary or physical link between these
two subsamples is likely to exist.

As a last point, we want to address the connection (if any) of ILLBGs
with the DRGs (Franx et al. 2003). Foerster-Schreiber et al. (2004)
and  Labbe et al. (2005) compared the properties of DRGs and blue LBGs
and found  that the former are, not surprisingly, on average dustier
and more massive.  When examining the M$/$L$_{k}$ mass-to-light ratio
they find that DRGs display  higher values (the mass-to-light ratios
remains of course critically  dependent on the assumed star formation
history, metallicity and IMF values) than the LBGs. The mean age 
and extinction of DRGs is $<$t$>$=1.3 Gyr and $<$A$_{V} >$1.5 mag.
From the SED fits
in Figure 2 of Labbe et al. (2005) we conclude that ILLBGs resemble
more the ``old and dusty'' DRGs rather than  the ``dead'' DRGs.

\section{\bf Massive LBGs: how many are there?}

Recently, it has become apparent that the number densities of
high-redshift galaxies with large stellar masses exceed the
theoretical predictions of semi-analytical models of galaxy evolution
(e.g. Saracco et al. 2004, Daddi et al. 2004, Tecza et al. 2004).  In
the widely accepted hierarchical merging scenario for galaxy
formation, massive galaxies are the end points of mergers which
increase with cosmic time.  In the semi-analytical models of Kauffmann
et al (1999),  the predicted number density of $\mathcal{M}_{\rm
star}\ge10^{11}$ $M_\odot $ ellipticals at $z\simeq2.5$ is about
$5-6\times10^{-5}$ Mpc$^{-3}$. A similar conclusion is reached by
Moustakas \& Somerville (2002). The semi-analytic models of galaxy
formation in the hierarchical clustering scenario by Baugh et
al. (1998, 2002) predict, in fact, no such massive galaxies at
redshift z$>$2. 

In Figure 11 we plot the evolution of the number density  $\it {\Phi}
(M > 10^{11} M_{\odot})$ based on the massive LBGs (assuming  an
$\Omega_m=0.3$, $\Omega_\Lambda=0.7$ cosmology with H$_{0}$=70 km s$^{-1}$
Mpc$^{-1}$) as a function of redshift. If we assume 
an effective volume for the U-dropouts 
V = 450 h$^{-3}$ Mpc$^{3}$ arcmin$^{-2}$
(taken from from Steidel et al. 1999) then, the effective co-moving 
volume becomes
V = 1400 Mpc$^{3}$ arcmin$^{-2}$ (the volumes have been weighted according to 
the number of objects per R-magnitude bin).
For the 8 micron
LBGs with M $>$10$^{11}$ M$_{\odot}$ in the 227 sq. arcmin Westphal field, 
we derive a co-moving density
of $\Phi =(1.6\pm0.5)\times 10^{-5}$ Mpc$^{-3}$ at the average redshift
$\langle z\rangle\simeq3$. The derived number density is an actual 
underestimate of the real number density of galaxies with $\mathcal{M}_{\rm
star}\ga10^{11}$ $M_\odot $ at $z>3$ since our calculation refers only to
the 8 micron selected massive LBGs. Although we are fairly confident that the
8 micron LBG sample contains {\it all} massive LBGs it does not, of course, 
account for the optically faint massive population at z$\sim$3. However, 
deriving $\Phi$  for other $z>3$ objects (e.g. the DRGs of Franx et al.) 
is a difficult task due to the very limited availability of measured 
spectroscopic redshifts for such targets. To date LBGs
consitute by far the largest $z \sim 3$ sample with measured spectroscopic 
redshifts and thus we confined our calculations of the number density to
this sample.

The current value of $\Phi$ 
is about a factor 3 higher than the predictions of hierarchical models by
e.g. Kauffmann et al. (1999). In Figure 11 we also plot
the co-moving densities of galaxies at z$\sim$0 (Cole et al. 2001), 
z$\sim$1 (Drory et al. 2004), z$\sim$2 (Fontana et al. 2004). 
In the same plot we show the theoretical predictions from semianalytical models
of Kauffmann et al. (1999), and Baugh et al. (2003). 
It is evident from Figure 11 that the evolution of the
number density of massive ( $\mathcal{M}_{\rm stars}\simeq10^{11}$
$M_\odot $) galaxies with redshift is slower than the prediction of 
the current hierarchical models, at least in the redshift range
0$<$z$<$3. We note, however, that the number densities shown 
in Figure 11 are based on a variety of surveys with different selection 
techniques. This, undoubtedly, introduces selection biases. Although 
disentagling the various biases is beyond the scope of the present work, we 
stress that despite the selection effects the main conclusion regarding the 
number density $\Phi$ of z$\sim$3 galaxies remains unaffected.

The fact that these galaxies have already assembled this mass at z$\sim$3, 
places the possible merging event of their formation at z$\geq$3.5 assuming a
dynamical time scale of 3$\times$10$^{8}$ yr (e.g. Mihos \& Hernquist 1996).
The inferred mass weighted age of the stellar populations places the formation
of their bulk at z$_{f} \geq$ 3.5 implying substantial activity at such high 
redshifts. Ferguson et al. (2002) postulated that the assembly of massive 
systems at z$\sim$3 can happen if star formation in LBGs is episodic 
and proceeds with a top heavy IMF. Another possibility is that the bulk of the
stellar mass was formed during a single star formation episode with large
$\tau$ (modeled as a single burst with large $\tau$ resembling conditions 
similar to a CSF model).

\section{\bf{Conclusions}}

The addition of long wavelength Spitzer$/$IRAC data
reveals, for the first time, that LBGs are rather inhomogeneous in their
rest--frame near-infrared properties. While very little scatter is observed  
in the optical (rest--frame UV properties), IRAC has revealed the
existence of a distinct class of (rest--frame) near--infrared luminous 
LBGs whose properties deviate from those of typical blue less massive LBGs.

Using the BC03 stellar population synthesis code we have generated
simple  models with exponentially decaying and$/$or constant star
formation histories to fit the LBG SEDs  and estimate stellar
masses. While in most cases both star formation models provide a good
fit to the LBG SEDs, there is a specific sub-group of LBGs (luminous
at 8 microns) that are best fit by a constant star formation history.
This subgroup of LBGs are also luminous in the mid--IR and are
detected by MIPS 24 micron channel. Incidentally, these are among the
most massive z$\sim$3 LBGs with estimated stellar masses in excess of
10$^{11}$M$_{\odot}$.

We have compared the properties of the 8 micron LBG sample to the near-IR
selected sample of Shapley et al. (2001). The 8 micron LBGs
form a slightly different subclass of ``redder'' LBGs. The results of their
SED modelling reveals that 8 micron LBGs are dustier 
(with marginally older ages) and also have higher masses. The need for a higher
dust extinction is only apparent when the IRAC data are included in the fit.

While the addition of the IRAC data has improved on the accuracy of
the estimated stellar masses the biggest gain has come from the tight
correlation between the absolute [8.0] micron magnitude and the
infrerred stellar mass.  The [8.0] micron magnitude
(rest--frame K--band) correlates well with the stellar mass estimates
and that in turns implies that the observed mid--infrared absolute
magnitudes ``trace'' the stellar mass much better than the optical magnitudes.
We infer an average  M$_{*} /$L of $\sim$ 0.2 although the values
for the most massive galaxies show a spread of about 10.
It is interesting to note that the massive LBGs also display 
``red colors'' (index R--[3.6]) similar to those found by 
Wilson et al. (2004) and to the colors of submm-galaxies 
(e.g. Egami et al. 2004). 

With such high masses and star formation rates LBGs along with submm galaxies
could well be the progenitors of today's ellipticals. It is thus natural to
investigate the connection between the two population from a mid--infrared
point of view. The connection between LBGs and submm galaxies has been
addressed indirectly by Chapman et al. (2005). However, it is only with the
availability of the IRAC observations (See Figure 1) that a proper comparison
can be made. Huang et al. (2005) have addressed this issue at length.
Given the ever increasing number of high redshift galaxies with high
masses (stellar or dynamical)
it is of interest to investigate the number density for these massive, 
high-redshift systems.
In the semi-analytical rendition of most hierarchical models for structure 
formation the predicted number density of massive M$_{*} >10^{11}M_{\odot}$
galaxies decreases with redshift. The number density we derived based on the
8 micron LBG sample (with M$_{*} >10^{11}M_{\odot}$) is
$\Phi =(1.6\pm0.7)\times 10^{-4}$ Mpc$^{-3}$ is at least a factor of 3 higher
than semi-analytical predictions. If we compare the density of local 
L$\geq$L$_{*}$ galaxies with our estimate we find that their density cannot
decrease by more than a factor of 3 from z=0 to z=3 thus we conclude that
a significant fraction of the stellar mass in local massive galaxies 
must have been in place at z$\sim$3.

\acknowledgements

We thank the anonymous referee for his/her suggestions on improving 
the manuscript. We thank S. Miyazaki for making the SUBARU R-band image
available. 
This work is based on observations made with the Spitzer Space
Telescope, which is operated by the Jet Propulsion Laboratory,
California Institute of Technology under NASA contract 1407. Support
for this work was provided by NASA through Contract Number 1256790
issued by JPL. DR acknowledges support from the Leverhulme Trust via 
a Research Fellowship grant.

{}
\clearpage

\begin{deluxetable}{rrrrrrr}
\tablecolumns{7}
\tablewidth{0pc}
\tablenum{1}
\footnotesize
\tablecaption{Ground based data for the 8 micron LBG sample with z$_{spec}$}
\tablehead{
\colhead{Name$^{1}$}&
\colhead{U$^{2,3}$}& 
\colhead{G$^{2,3}$}&
\colhead{R$^{2,3}$}&
\colhead{J$^{2,3}$}&
\colhead{K$^{2,3}$}&
\colhead{z$_{spec}^{4}$}}
\startdata
C10& 28.22& 25.55& 25.33& & & 3.053\\
M30& 28.09& 25.62& 24.57&  && 3.380\\
MD9& 26.65 & 25.18 & 24.9 && &3.035\\
D29& 27.43 & 25.14 & 24.82 & 24.12 & 23.78 &3.245\\ 
D49&25.95 & 23.67 & 23.5 & 23.07 & 21.976& 2.808\\
M63& 27.8& 25.78& 24.82& & &3.101\\
C49& 28.03& 25.7& 24.93& & &  2.932\\
C69& 27.68& 25.08& 24.24& & & 2.948\\
C15& 27.84& 25.13& 24.07& & & 3.024\\
C47&27.67& 25.02& 24.44& & &2.973\\
C50& 27.63 &24.63& 23.96& & & 2.910\\ 
C58& 27.97& 25.00& 24.51& & & 2.747\\
C76&27.54& 24.12& 23.33& & &  2.876\\
D40&27.5& 24.44& 23.67& & & 2.958\\ 
D55&26.54& 24.1& 23.66&  & & 2.994\\
M28&27.42& 25.49& 24.65& & & 2.903\\ 
M38& 27.9& 25.75& 24.86& & & 2.928\\
M41&28.14& 26.09& 25.4& & &3.396\\ 
M43&27.96& 25.77& 24.78& & &3.351\\
MD23&26.91& 25.04& 24.22&  & 23.23& 2.862\\
MD91&26.23& 24.33& 23.84& 23.63& 22.93& 2.740\\ 
MD97& 25.3& 23.81& 23.43& & & 2.496\\ 
MD98& 27.73& 25.57& 24.64& & & 3.119\\
MD99&27.55& 25.06& 24.02& & & 3.363\\ 
MD109& 26.62& 24.78& 23.94&  & 22.71& 2.719\\ 
MD115& 26.27 &24.62& 23.97&  & 24.17& 3.208\\
\enddata
\tablenotetext{1}{The naming convention is the same as in Steidel et al. 2003, but we have omitted the Westphal prefix.}
\tablenotetext{2}{All magnitudes are in AB units.}
\tablenotetext{3}{Magnitudes taken from Steidel et al. (2003)}
\tablenotetext{4}{All confirmed spectroscopic redshifts are from
emission or absorption features, Steidel et al. (2003)}
\end{deluxetable}
\begin{deluxetable}{rrrrrrrrrrrr}
\tablecolumns{12}
\tablewidth{0pc}
\tablenum{2}
\footnotesize
\tablecaption{Spitzer Properties of 8 micron LBG sample with $z_spec$}
\tablehead{
\colhead{Name$^{1}$}&
\colhead{[3.6]$^{2}$}&
\colhead{[4.5]$^{2}$}&
\colhead{[5.8]$^{2}$}&
\colhead{[8.0]$^{2}$}&
\colhead{[24]$^{2}$}}
\startdata
C10& 24.83$\pm$0.31& 24.68$\pm$0.39& & 23.95$\pm$0.12&  \\
M30& 23.04$\pm$0.14& 22.67$\pm$0.12&   & 22.19$\pm$0.12& 19.35$\pm$0.45\\
MD9&  22.40$\pm$0.07& 22.30$\pm$0.09& 22.42$\pm$0.31& 22.77$\pm$0.12&  & \\
D29&  23.49$\pm$0.20& 23.08$\pm$0.16&  & 21.75$\pm$0.12&   \\ 
D49& 21.03$\pm$0.03& 20.72$\pm$0.05& 20.53$\pm$0.07& 20.39$\pm$0.06& 18.53$\pm$0.46&\\
M63& 22.79$\pm$0.11& 22.79$\pm$0.12&   & 22.69$\pm$0.12&  \\
C49&  22.33$\pm$0.07& 22.16$\pm$0.08&   & 21.78$\pm$0.12&  \\
C69&  22.68$\pm$0.10&22.63$\pm$0.11&  & 22.63$\pm$0.12& \\
C15&  22.78$\pm$0.11& 22.97$\pm$0.14&   & 22.13$\pm$0.12& \\
C47& 22.79$\pm$0.11& 22.68$\pm$0.11&   & 22.48$\pm$0.12&  \\
C50& 20.90$\pm$0.03& 20.62$\pm$0.02& 20.97$\pm$0.11& 20.49$\pm$0.07& 20.98$\pm$0.65\\ 
C58&  21.88$\pm$0.06& 21.79$\pm$0.05& 22.05$\pm$0.27& 21.52$\pm$0.12& 18.95$\pm$0.06\\
C76 & 22.49$\pm$0.12&  \\
D40& 22.13$\pm$0.06& 21.99$\pm$0.07& 22.07$\pm$0.27& 21.90$\pm$0.12&\\
D55& 22.81$\pm$0.12& 22.71$\pm$0.12& & 22.98$\pm$0.12&\\
M28& 20.50$\pm$0.02& 20.22$\pm$0.01& 20.51$\pm$0.07& 20.41$\pm$0.06& 18.24$\pm$0.30\\ 
M38& 23.40$\pm$0.11& 23.54$\pm$0.22&   & 22.70$\pm$0.12& \\
M41& 22.55$\pm$0.09& 22.43$\pm$0.10& 22.54$\pm$0.31& 22.23$\pm$0.12& \\ 
M43& 23.70$\pm$0.21& 23.36$\pm$0.19&   & 26.56$\pm$0.12& \\
MD23&22.45$\pm$0.08& 22.52$\pm$0.11& 21.92$\pm$0.14& 22.74$\pm$0.12&\\
MD91&22.63$\pm$0.09&22.44$\pm$0.10&   & 22.31$\pm$0.12&  \\ 
MD97&  21.59$\pm$0.05& 21.50$\pm$0.04& 21.07$\pm$0.11& 21.29$\pm$0.12& \\ 
MD98& 23.10$\pm$0.15& 23.09$\pm$0.16&   & 22.67$\pm$0.12&  \\
MD99& 21.64$\pm$0.05& 21.41$\pm$0.04& 
21.21$\pm$0.12& 20.72$\pm$0.08& 17.39$\pm$0.09\\ 
MD109& 22.55$\pm$0.09& 22.52$\pm$0.11&  & 22.68$\pm$0.12&  \\ 
MD115& &23.53$\pm$0.23&  &22.80$\pm$0.12&  \\
\enddata
\tablenotetext{1}{The naming convention is the same as in Steidel et al. 2003, but we have omitted the Westphal prefix}
\tablenotetext{2}{All magnitudes are in AB units}
\end{deluxetable}

\begin{deluxetable}{rrrrr}
\tablecolumns{5}
\tablewidth{0pc}
\tablenum{3}
\footnotesize
\tablecaption{Modelling results for the 8 micron LBG sample}
\tablehead{
\colhead{Name}&
\colhead{E(B--V)}& 
\colhead{t$_{sf}^{1}$}&
\colhead{log M$_{*}$(CSF)$^{2}$}&
\colhead{log M$_{*}$(EXP)$^{3}$}}
\startdata
C10&0.214& 65& 9.56$\pm$0.11& 9.82$\pm$0.09\\
M30& 0.201& 2076& 10.83$\pm$0.18& 10.91$\pm$0.22\\
MD9& 0.123& 1542& 10.48$\pm$0.05& 10.62$\pm$0.09\\
D29& 0.184& 402& 10.45$\pm$0.11& 10.37$\pm$0.14\\
D49& 0.340& 1350& 11.06$\pm$0.09& 11.10$\pm$0.06\\
M63& 0.104& 551& 10.71$\pm$0.13& 10.84$\pm$0.18\\
C49& 0.211& 351& 10.62$\pm$0.24& 10.71$\pm$0.19\\
C69& 0.111& 641& 10.53$\pm$0.08& 10.61$\pm$0.11\\
C15& 0.251& 850& 10.68$\pm$0.12& 10.84$\pm$0.14\\
C47& 0.187& 1250& 10.69$\pm$0.22& 10.78$\pm$0.13\\
C50& 0.223& 1009& 11.58$\pm$0.15& 11.69$\pm$0.08\\
C58& 0.224& 1351& 10.62$\pm$0.06& 10.85$\pm$0.09\\
C76&0.378& 346& 10.92$\pm$0.28& 10.78$\pm$0.22\\
D40& 0.118& 910& 10.41$\pm$0.07& 10.53$\pm$0.11\\
D55& 0.092& 989& 10.27$\pm$0.14& 10.32$\pm$0.11\\
M28&0.254& 1140 &11.48$\pm$0.13 &11.51$\pm$0.19\\
M38& 0.322& 48& 9.76$\pm$0.09& 9.84$\pm$0.05\\
M41& 0.100& 1933& 11.06$\pm$0.04& 11.12$\pm$0.08\\
M43& 0.258& 267& 10.69$\pm$0.16& 10.77$\pm$0.10\\
MD23&0.34& 96& 10.48$\pm$0.25& 10.56$\pm$-.19\\
MD91&0.224& 375& 10.49$\pm$0.13& 10.52$\pm$0.26\\
MD97& 0.331& 104& 10.46$\pm$0.07& 10.54$\pm$0.12\\
MD98& 0.208& 114& 10.46$\pm$0.11& 10.51$\pm$0.22\\
MD99& 0.264& 2112& 11.06$\pm$0.10& 11.12$\pm$0.14\\
MD109& 0.66& 67& 10.43$\pm$0.19& 10.67$\pm$0.11\\
MD115& 0.100& 1551& 10.48$\pm$0.09& 10.61$\pm$0.05\\
\enddata
\tablenotetext{1} {Age in Myrs, for constant star formation model.}
\tablenotetext{2} {Derived stellar mass and uncertainty in solar units, for constant star formation model.}
\tablenotetext{3} {Derived stellar mass and uncertainty in solar units, for
the best-fitting $\tau$ model. The stellar mass uncertainty reflects the uncertainty in $\tau$.}
\end{deluxetable}
\clearpage

\begin{figure}
\includegraphics[width=12cm,angle=-90]{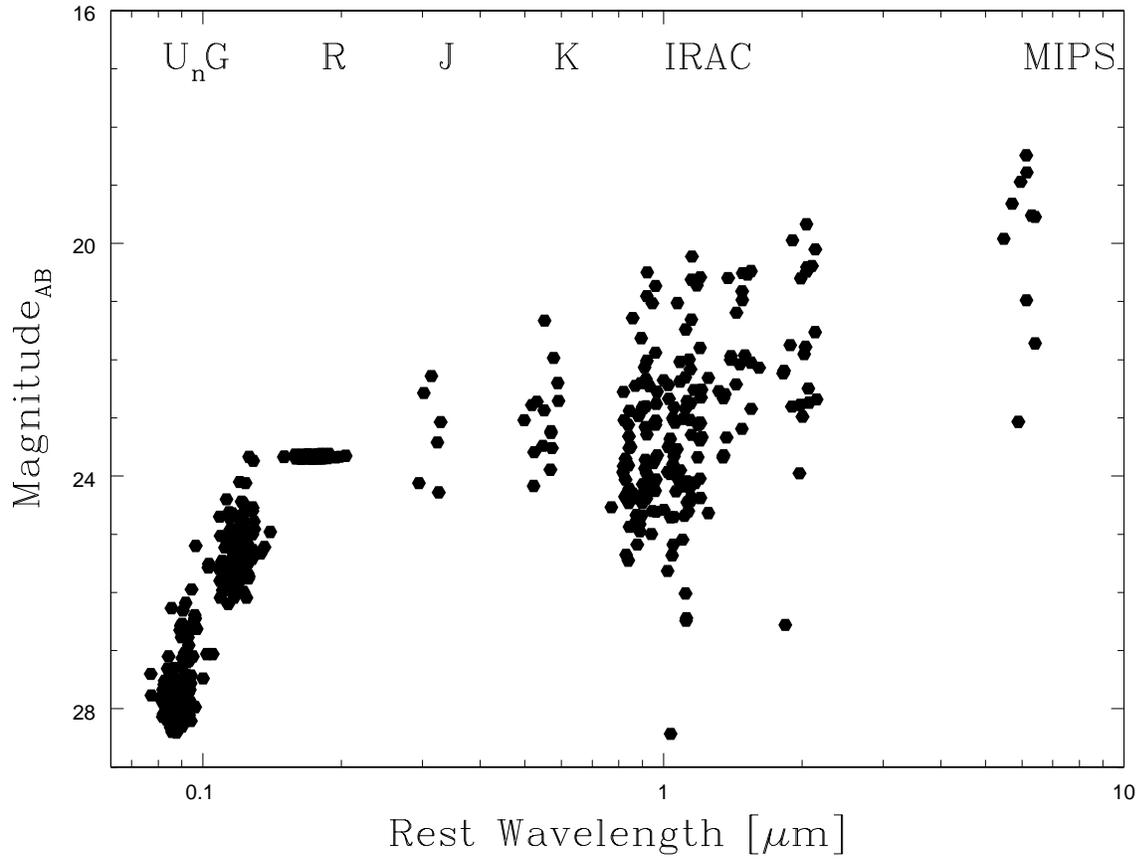}
\caption{The rest--frame UV$/$optical$/$near--IR SEDs of all LBGs 
detected in at least two Spitzer bands in the EGS area with spectroscopic 
redshifts. The UGR data are from the compilation of Steidel et al. (2003).
The SEDs are normalised in R. The U$_{n}$ magnitudes for the ``C'' LBGs are
upper limits.
\label{fig1}}
\end{figure}
\clearpage

\begin{figure}
\includegraphics[width=12cm,angle=-90]{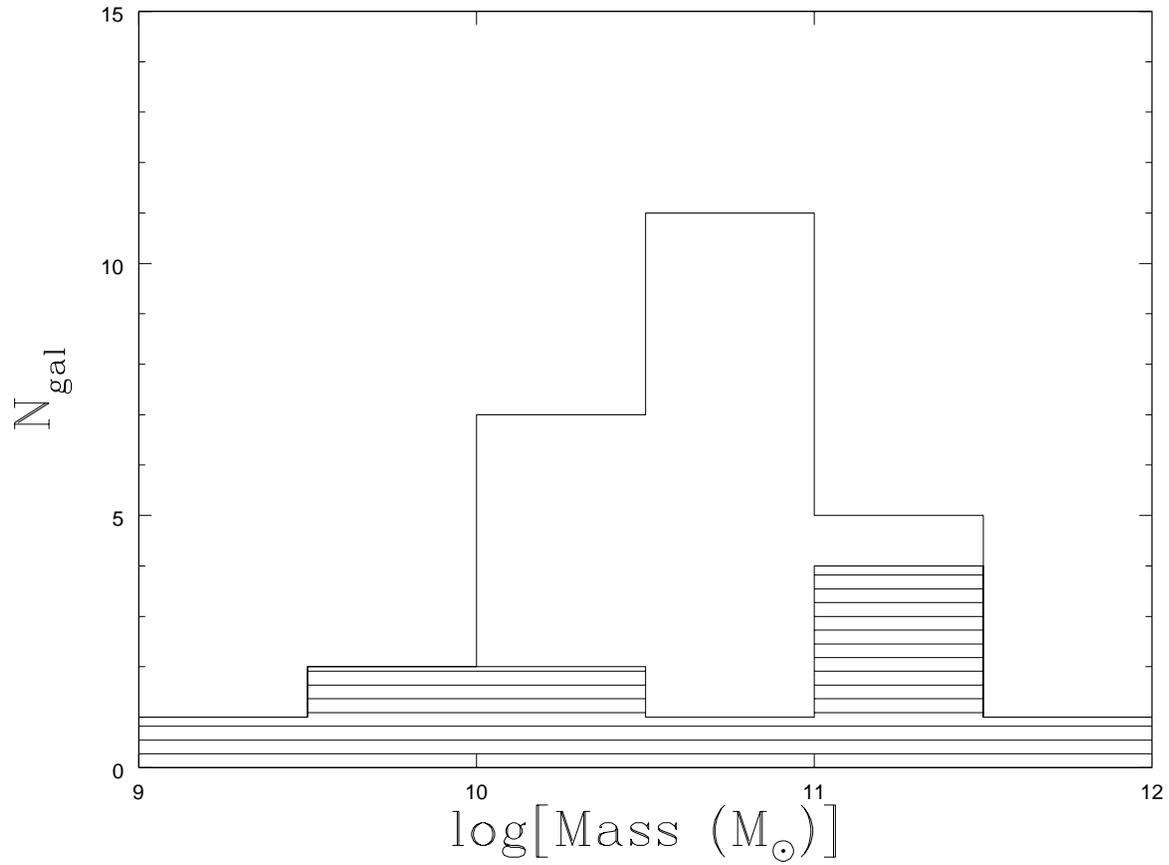}
\caption{Distribution of the stellar masses derived from the
best-fitting models for the entire sample of Spitzer LBGs with
spectroscopic redshifts. The models assume a Salpeter IMF (0.1 -- 100
M$_{\odot}$) solar metallicity and the Calzetti (2000) extinction
law. The shaded area refers to the distribution of the stellar masses
of the 8 micron LBG sample.}
\label{fig2}
\end{figure}

\begin{figure}
\epsscale{0.9}
\plotone{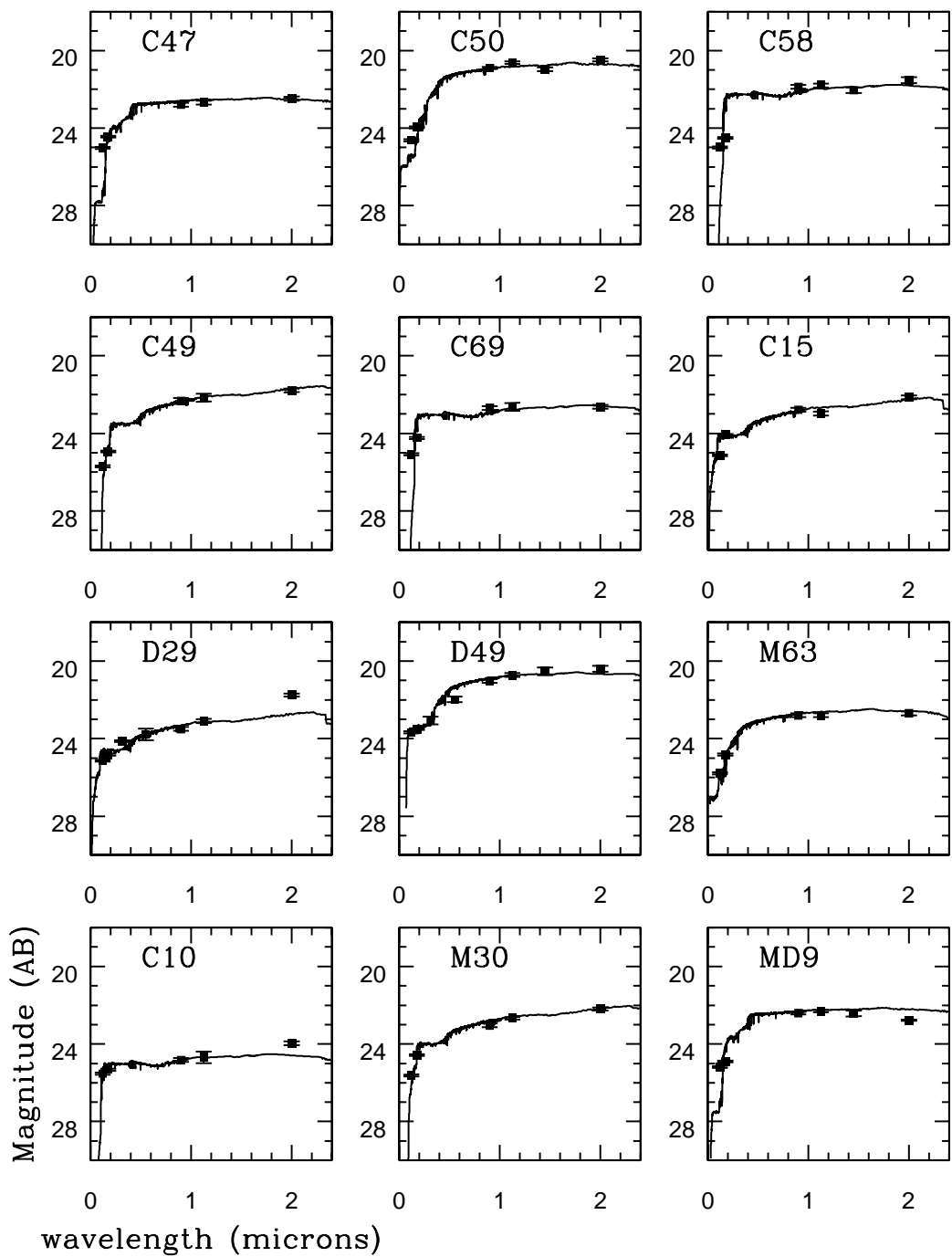}
\caption{ Best-fit BC03 constant star formation models for the 26 LBGs with
confirmed spectroscopic redshifts from the 8 micron LBG sample. The wavelength
axis is in the rest-frame. The solid lines  represent
the best fit BC 2003 model (either CSF: constant star formation or 
EXP: exponentially decaying star formation).
}
\end{figure}
\clearpage
\plotone{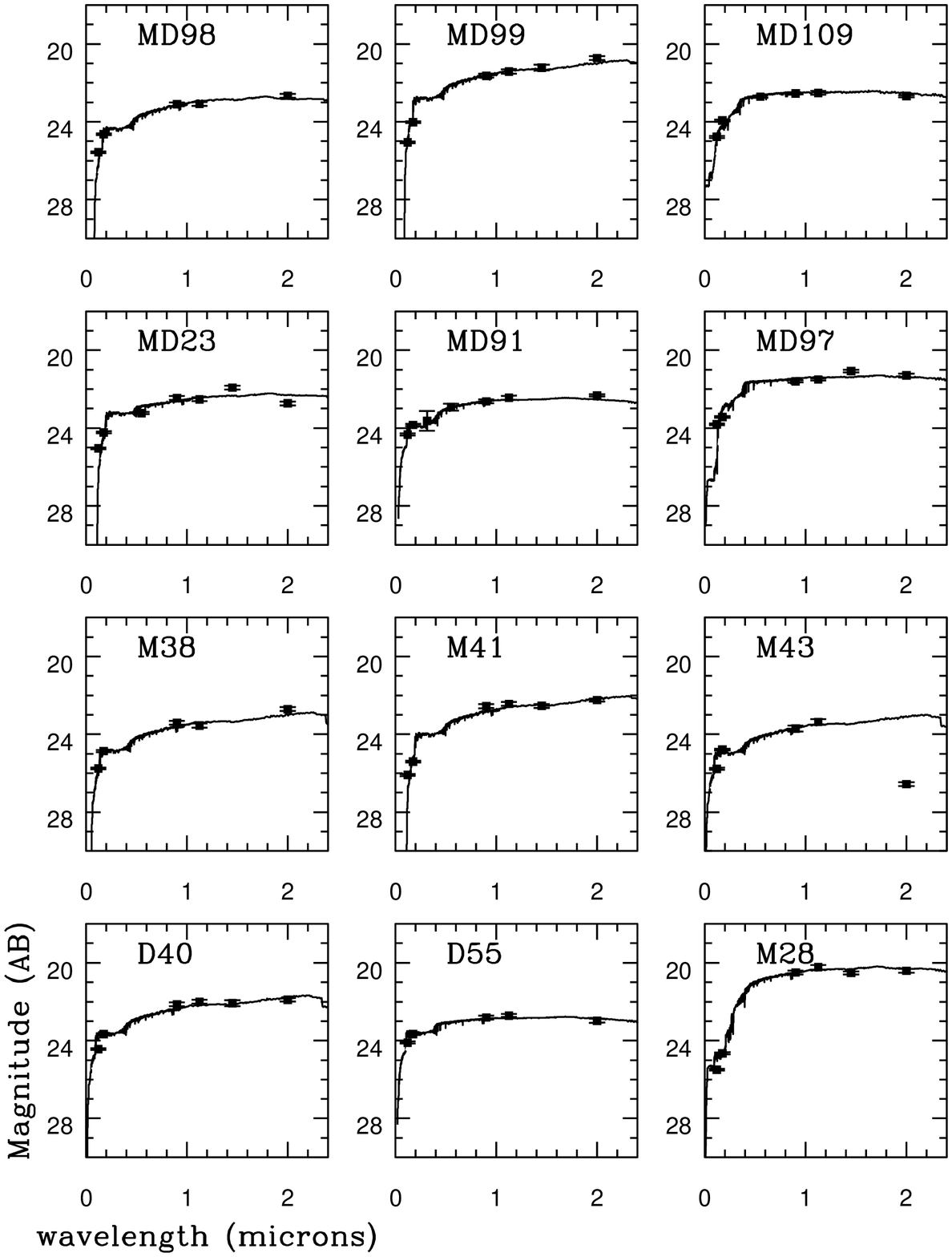}
\centerline{Fig. 3. --- Continued.}
\clearpage
\plotone{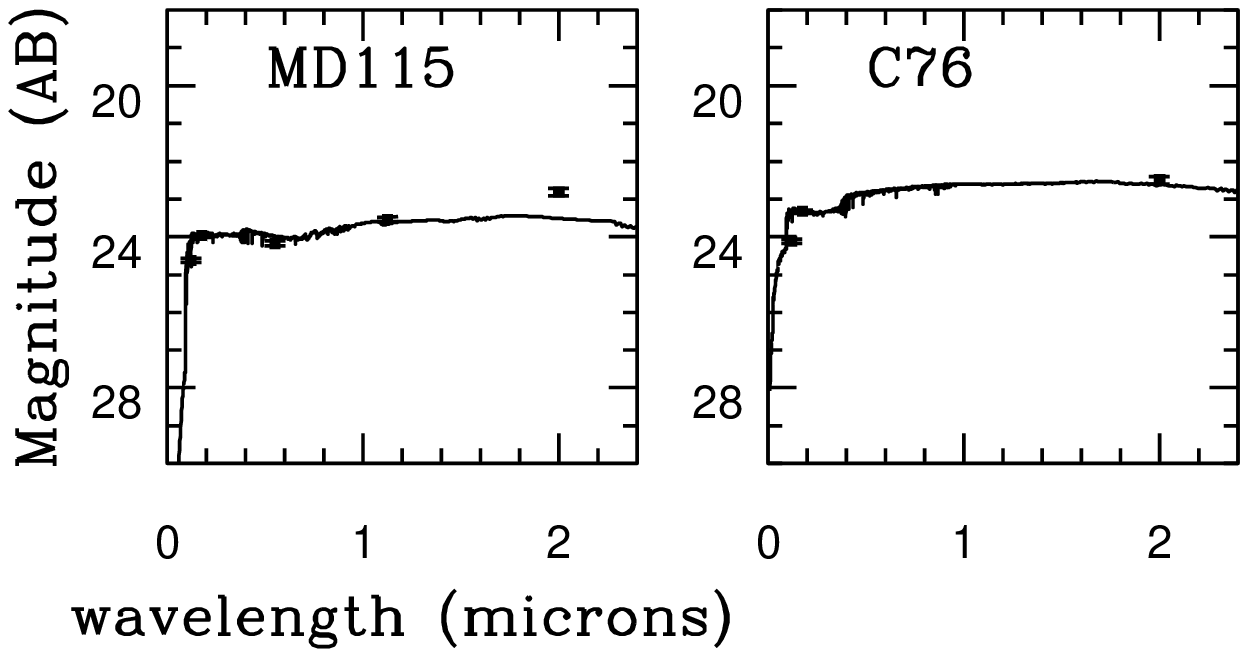}
\centerline{Fig. 3. --- Continued.}
\clearpage

\epsscale{1}
\begin{figure}
\begin{minipage}[t]{6cm}
\includegraphics[width=5cm, angle=-90]{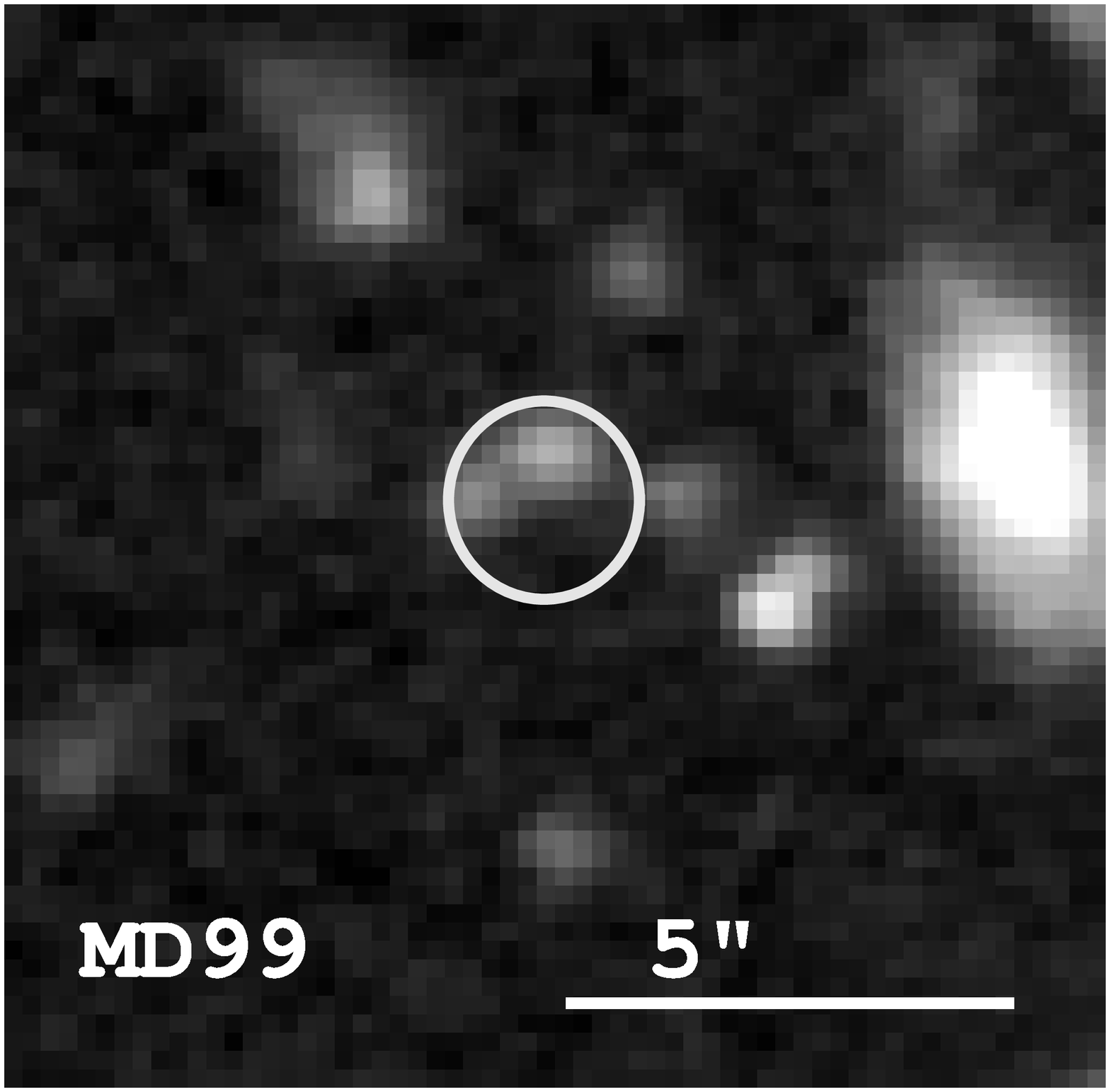}
\end{minipage} 
\hfill
\begin{minipage}[t]{6cm}
\includegraphics[width=5cm, angle=-90]{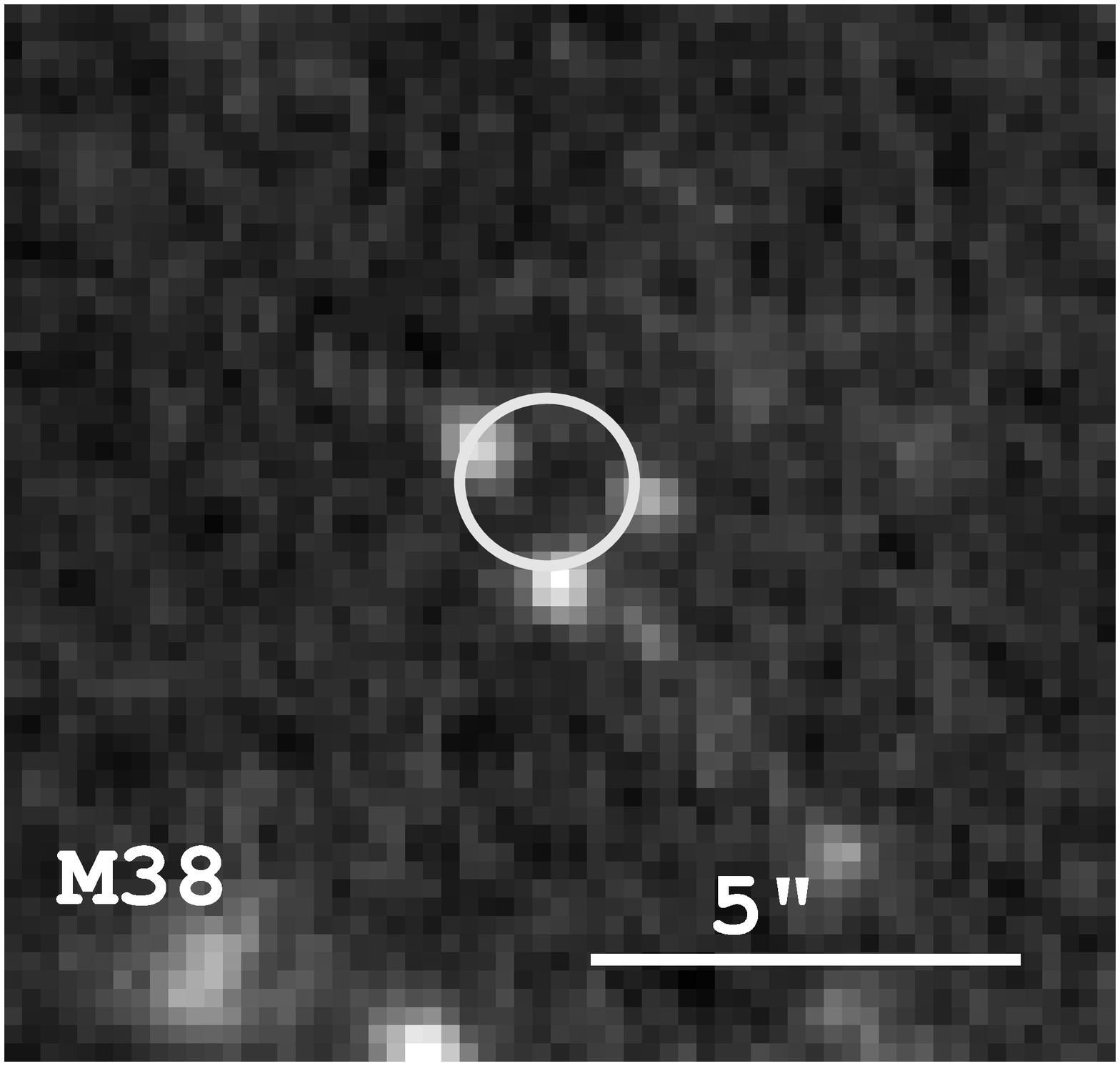}
\end{minipage}
\hfill
\caption{Subaru-R band postage stamps for two LBGs, Westphal-MD99 and 
Westphal-M38 where the 8 micron Spitzer
flux might be contaminated due to the complex morphology of the sources.
The images are 11$\arcsec\arcsec$ on a side and the circle indicates a
2 $\arcsec$ aperture.}
\label{fig4}
\end{figure}								
  						
\clearpage

\begin{figure}
\includegraphics[width=12cm,angle=-90]{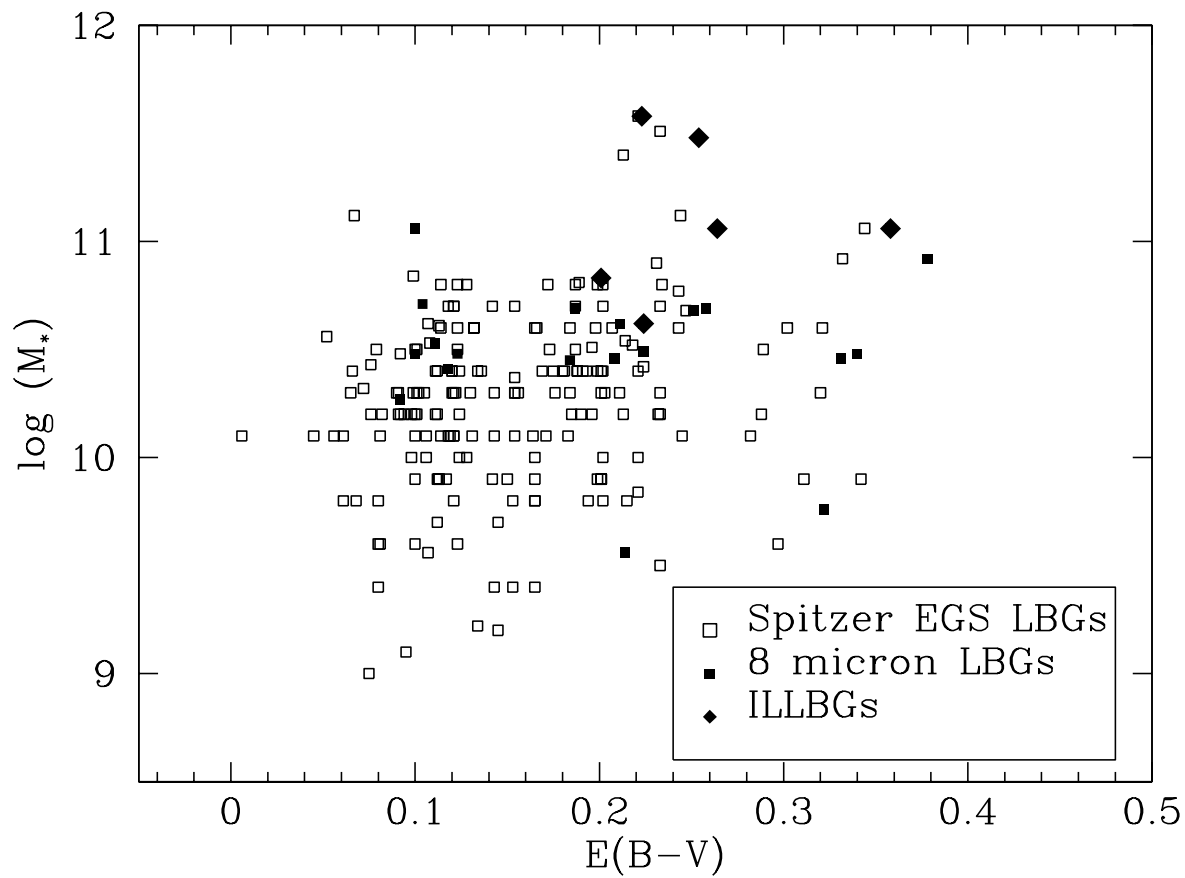}
\caption{Stellar mass as a function of extinction. Parameters are derived from
the best-fit BC03 models with constant star formation.  Open squares
denote all galaxies detected by Spitzer, filled squares denote the 38
non-ILLBG galaxies in the 8~\micron\ LBG sample, and large filled
diamonds denote the six ILLBGs, which are also in the 8~\micron\ LBG
sample. For clarity
we mark the ILLBGs (part of the 8 micron LBG sample) as asterisks.}
\label{fig5}
\end{figure}

\begin{figure}
\begin{centering}
\includegraphics[width=12cm,angle=-90]{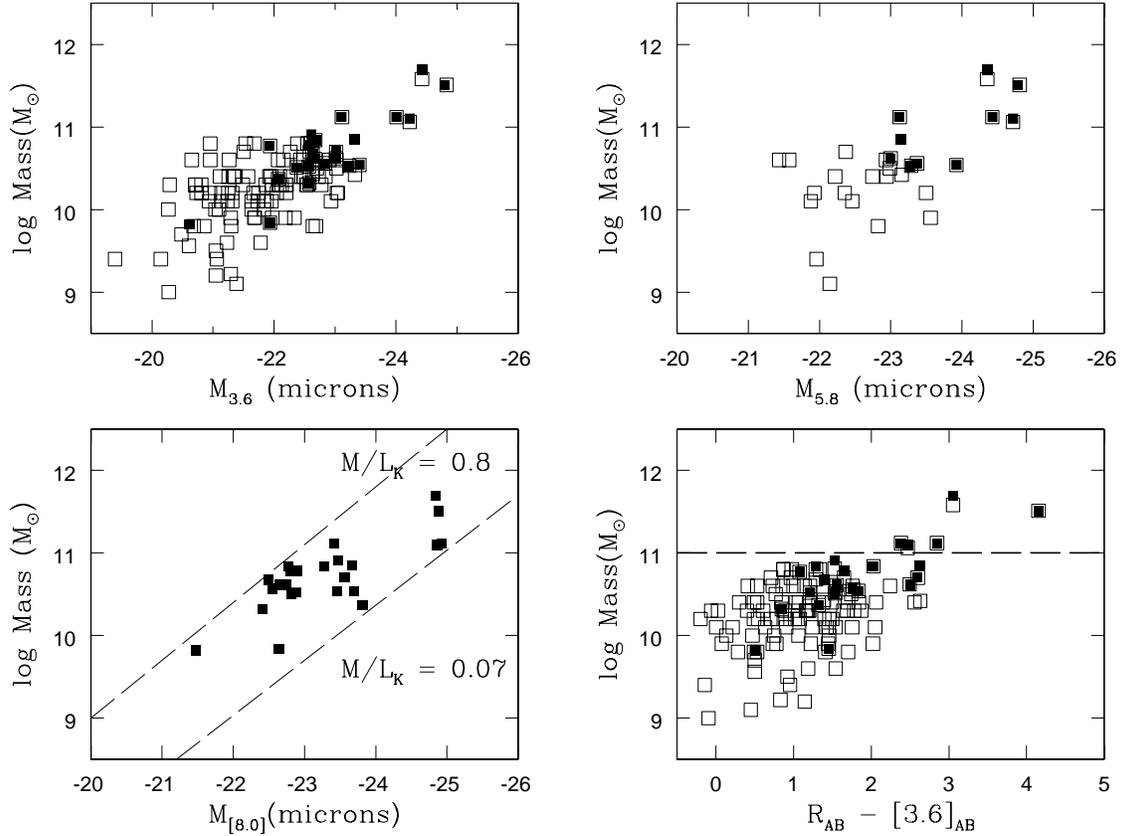}
\caption{Estimated stellar masses from the best fit 
model as a function of absolute [3.6] micron magnitude, rest--frame I-band 
(top left), absolute [5.8] micron magnitude, rest--frame H--band (top right),
for all LBGs with confirmed spectroscopic 
redshifts. The filled 
symbols correspond to the 8 micron LBGs (with known spectroscopic redshifts).
The lower left panel shows inferred  stellar masses as a function 
of absolute [8.0] micron 
magnitude (8 micron LBG sample). 
The scatter in stellar mass-to-light ratio at a given
[8.0] micron luminosity (rest--frame K band luminosity at redshift 3) 
can be as high as 12. The dashed lines indicate the range of stellar M$/$L 
in the sample and are given in solar units evaluated at rest--frame K-band.  
The lower right panel shows stellar masses as a function of R--[3.6] colour.
There is a strong corellation between masses and the 
R--[3.6] colour particularly for the 8 micron LBG sample (filled squares). 
The most massive objects (M$>$10$^{11}$M$_{\odot}$) tend to show the
reddest R--[3.6] colors as well.}
\label{fig6}
\end{centering}
\end{figure}

\begin{figure}
\includegraphics[width=12cm,angle=-90]{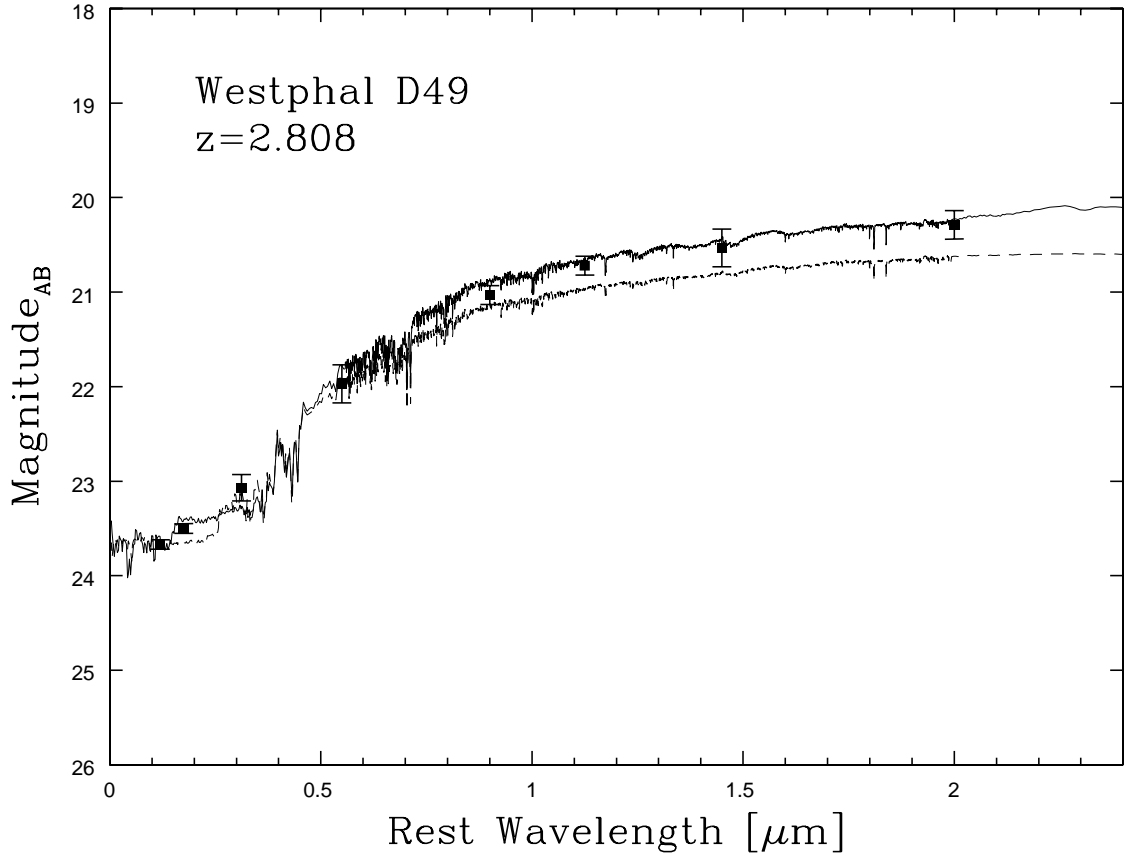}
\caption{A comparison of two BC03 models assuming constant star formation 
history but varying extinction. The dashed line shows a model with 
E(B-V)=0.17 and $t_{sf}$ = 1139 Myrs as suggested by Shapley et al. (2001) 
based on their analysis. The solid line shows our model, E(B-V)=0.34 and  
$t_{sf}$=1350 Myrs (fit including the IRAC photometry)}. 
\label{fig7}
\end{figure}

\begin{figure}
\includegraphics[width=12cm,angle=-90]{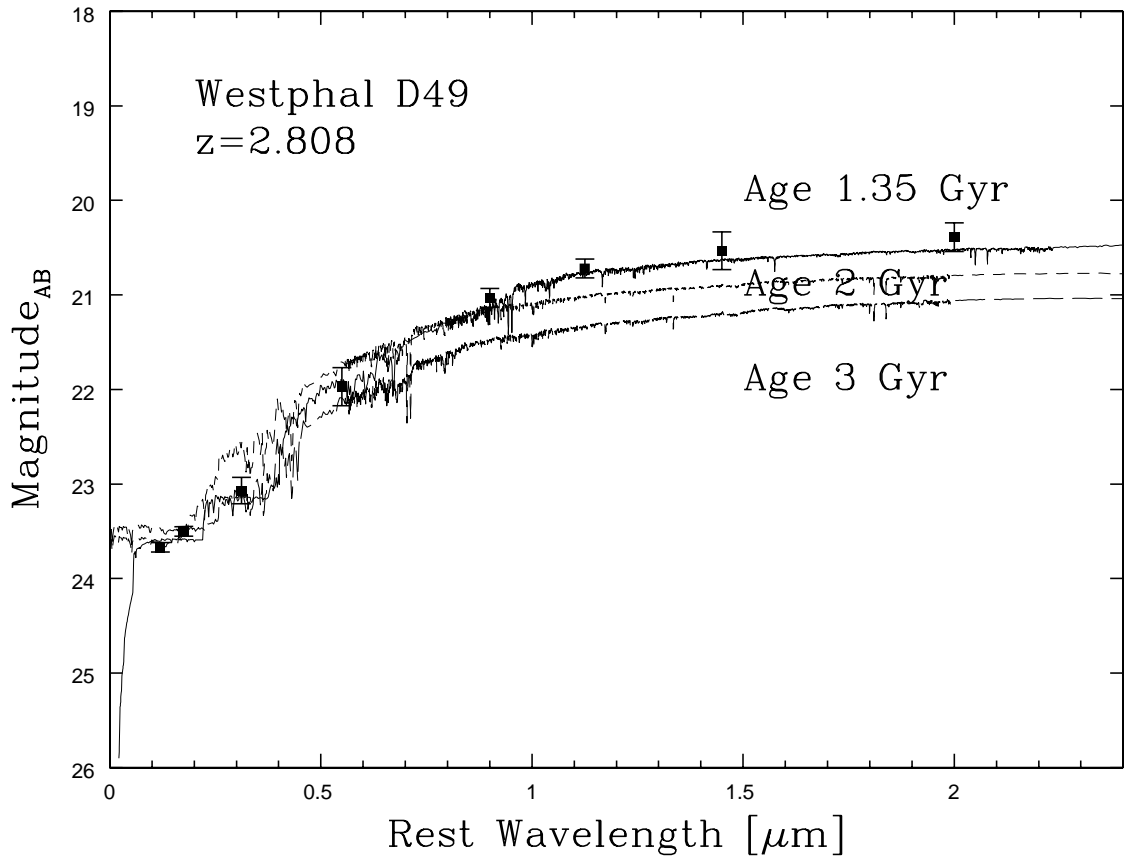}
\caption{A Comparison of three BC03 constant star formation models
with varying ages t$_{sf}$: 1.35 Gyrs (solid line), 
2.0 Gyrs (short dashed line) and 3 Gyrs (long dashed line). The comparison
(together with the results shown in Figure 7) demonstrates that the IRAC
photometry is best fit by a model with higher extinction rather
than older stellar ages.
}
\label{fig8}
\end{figure}

\begin{figure}
\includegraphics[width=12cm,angle=-90]{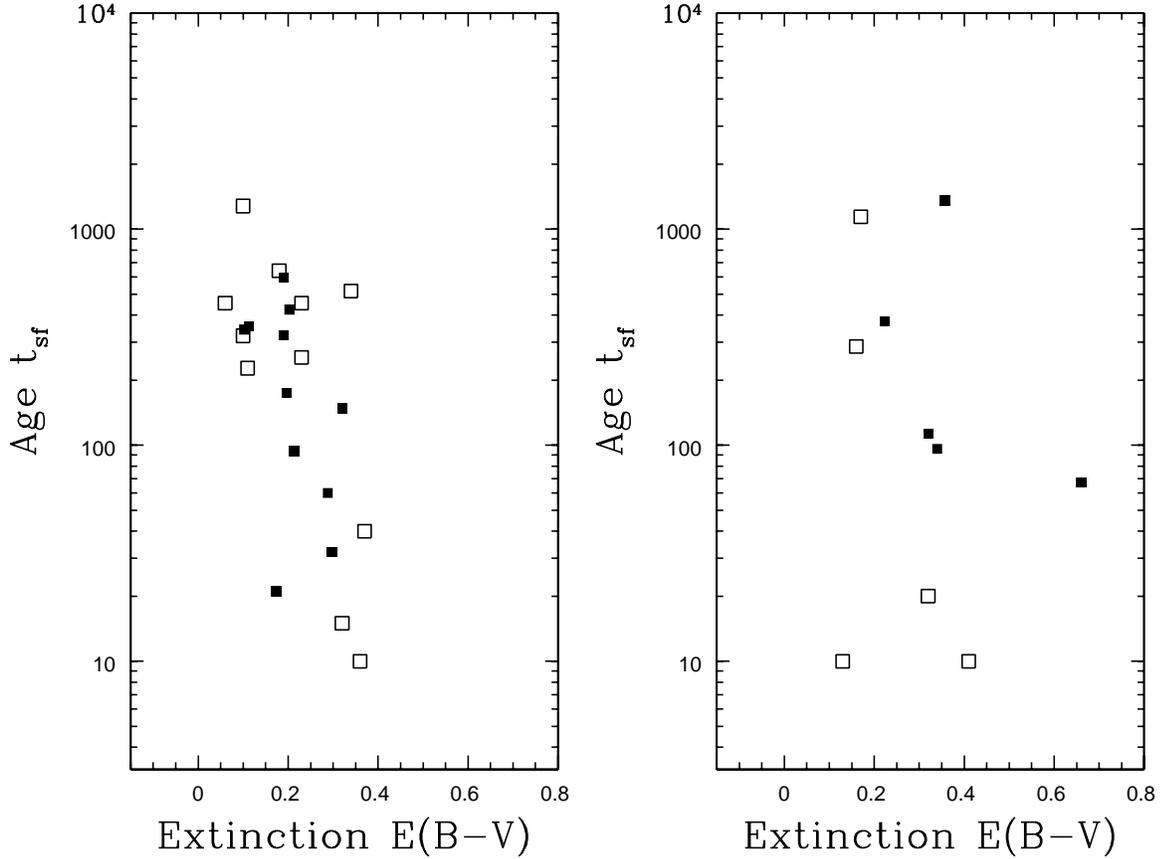}
\caption{
Age as a function of extinction. Sets of the best fit BC03 constant star 
formation extinction E(B--V) and age t$_{sf}$ (Myrs) parameters are 
plotted for:
(left panel) the 11 LBGs that are in common between the NIRC sample of 
Shapley et al. (open symbols ) and the Spitzer LBG sample (filled symbols). 
In the right panel
the same set of parameters (age and extinction) are plotted for the 5 NIRC LBGs
(open symbols) that are part of the 8 micron LBG sample (filled symbols).
The differences in the values of E(B--V) and t$_{sf}$ are more pronounced for
those LBGs with Spitzer 8 micron detections. Overall, the best fit E(B--V) 
values are higher for those LBGs with Spitzer 8 micron detections.}
\label{fig9}
\end{figure}

\begin{figure}
\includegraphics[width=16cm, angle=0]{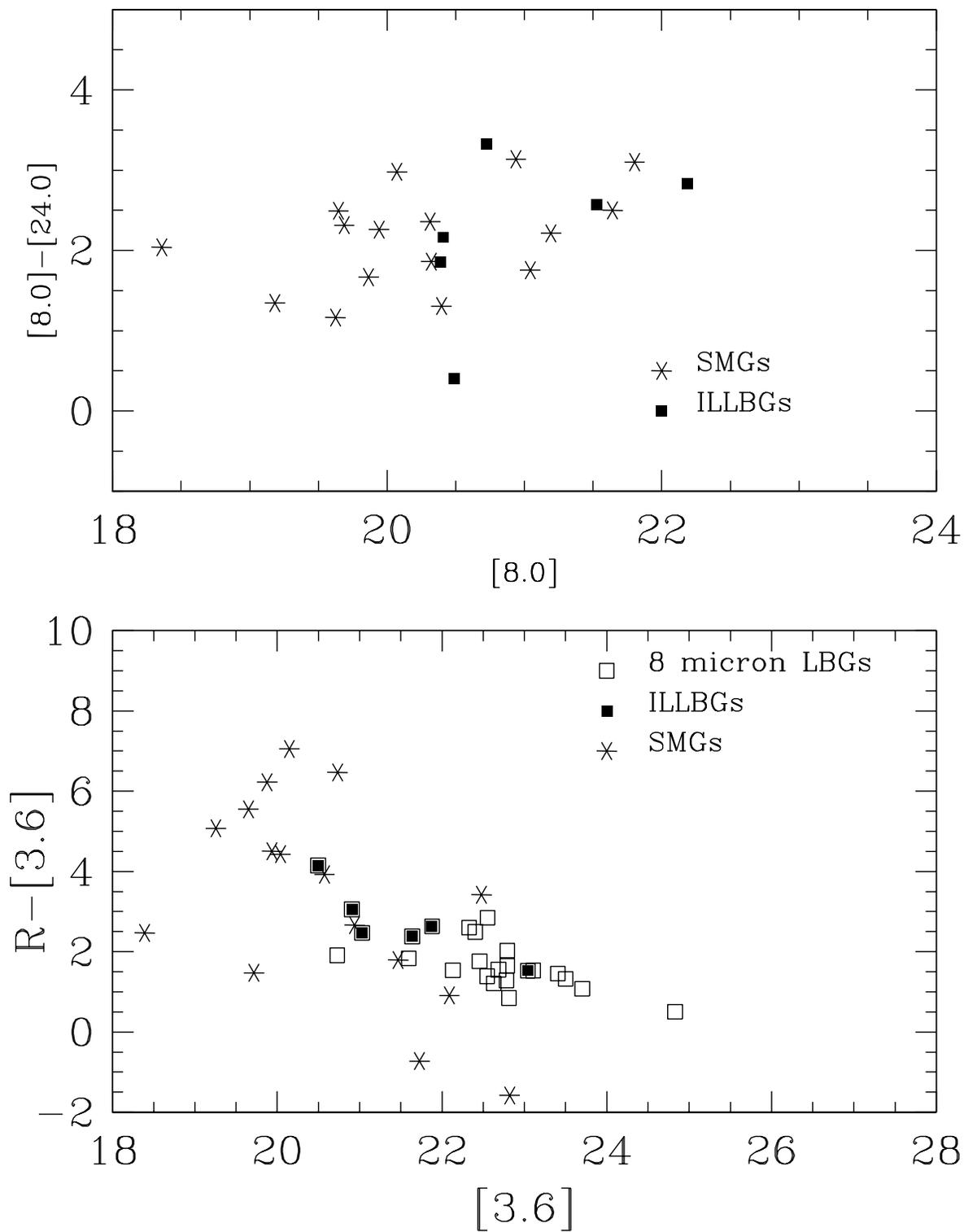}
\caption{(Lower panel): R-[3.6] vs. [3.6] color-magnitude plot for 8 
micron LBGs (open squares),
ILLBGs (filled squares) and SMGs (stars).
(Upper panel): [8.0]-[24.0] vs. [8.0] color-magnitude for ILLBGs 
(filled squares),
SMGs (asterisks). The data for the SMGs are from Ashby
et al. (2006). All magnitudes are in AB. }
\label{fig10}
\end{figure}

\begin{figure}
\includegraphics[width=12cm,angle=-90]{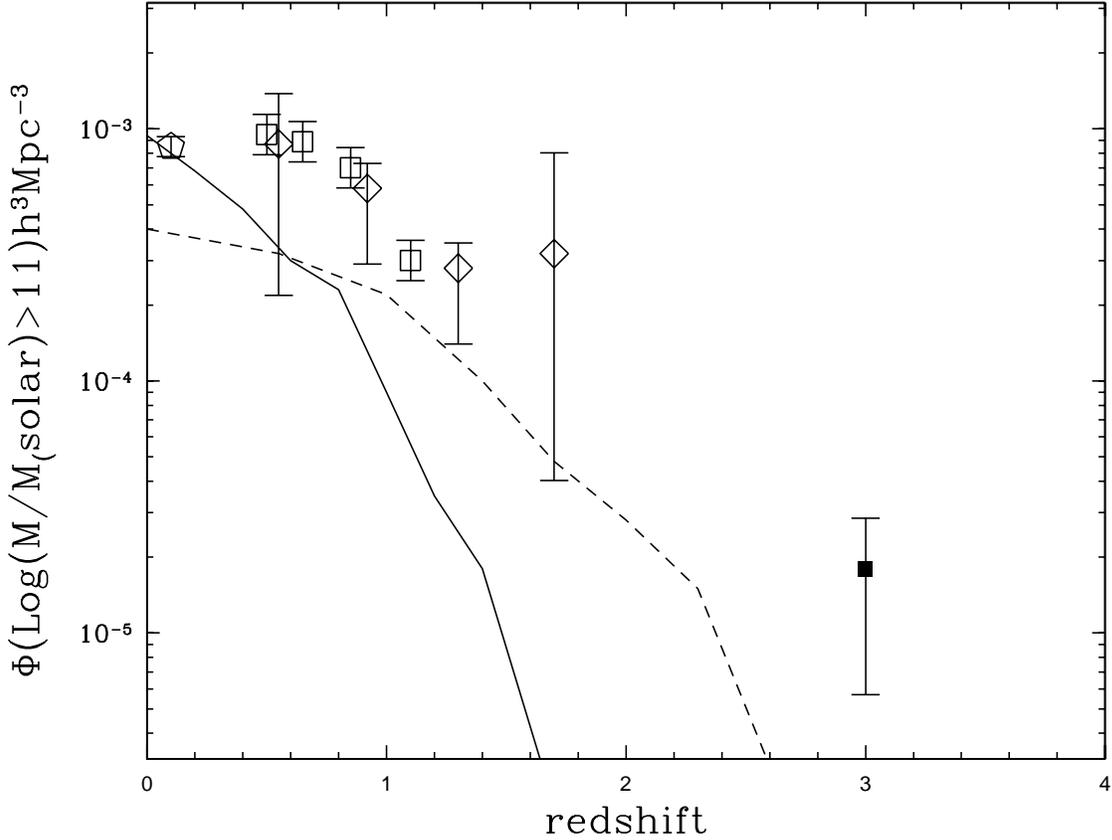}
\caption{Number density for galaxies with M$>$10$^{11}$M$_{\odot}$ as
a function of redshift (assuming a flat $\Omega_{\Lambda}$ = 0.7,
H$_{0}$ = 70 h$_{0.7}$ km s$^{-1}$ Mpc$^{-1}$). The filled square
represents the mass density of the 8 micron LBGs (with
M$>$10$^{11}$M$_{\odot}$). The open pentagon denotes the z=0 results
based on 2dF/2MASS data by Cole et al. (2000).  The open diamonds are
from Drory et al. (2004) and represent source densities for massive
early-type galaxies from the MUNICS survey. Open squares denote number
densities for massive red galaxies in the HDFS (Saracco et al. 2004).
 We show, for comparison, two theoretical predictions from
semi--analytical models: thick solid curve comes from the predictions
of Kauffmann et al. (1999), while dashed line represents the
predictions from Baugh et al. (2003).
\label{fig11}}
\end{figure}

\end{document}